\documentclass[%
 reprint,
superscriptaddress,
 amsmath,amssymb,
 aps,
prb,
floatfix,
]{revtex4-2}

\usepackage{graphicx}% Include figure files
\usepackage{dcolumn}% Align table columns on decimal point
\usepackage{bm}% bold math
\usepackage[version=4]{mhchem}
\newcommand{\basb}{\ce{BaTi2Sb2O}}
\newcommand{\babi}{\ce{BaTi2Bi2O}}

\newcommand{\bana}{\ce{Ba$_{1-x}$Na$_x$Ti2Sb2O}}

\newcommand{\bak}{\ce{Ba$_{1-x}$K$_x$Ti2Sb2O}}
\newcommand{\barb}{\ce{Ba$_{1-x}$Rb$_x$Ti2Sb2O}}
\newcommand{\bacs}{\ce{Ba$_{1-x}$Cs$_x$Ti2Sb2O}}
\newcommand{\assb}{\ce{BaTi2(As$_{1-x}$Sb$_x$)2O}}

\newcommand{\naas}{\ce{Na2Ti2As2O}}
\newcommand{\nasb}{\ce{Na2Ti2Sb2O}}
\newcommand{\tc}{$T_{\rm c}$}
\newcommand{\nef}{$N(E_{\rm F})$}

\begin{document}

\title{Theory for doping trends in titanium oxypnictide superconductors}

\author{Han-Xiang Xu}
\affiliation{Research Institute for Interdisciplinary Science, Okayama University, Okayama 700-8530, Japan}
\author{Daniel Guterding}
\affiliation{Fachbereich Mathematik, Naturwissenschaften und Datenverarbeitung, Technische Hochschule Mittelhessen, Wilhelm-Leuschner-Stra{\ss}e 13, 61169 Friedberg, Germany}
\author{Harald O. Jeschke}
\affiliation{Research Institute for Interdisciplinary Science, Okayama University, Okayama 700-8530, Japan}

\date{\today}

\begin{abstract}
A family of titanium oxypnictide materials \ce{BaTi2$Pn$2O} ($Pn = \text{pnictogen}$) becomes superconducting when a charge and/or spin density wave is suppressed. With hole doping, isovalent doping and pressure, a whole range of tuning parameters is available. We investigate how charge doping controls the superconducting transition temperature {\tc}. To this end, we use experimental crystal structure data to determine the electronic structure and Fermi surface evolution along the doping path. We show that a naive approach to calculating {\tc} via the density of states at the Fermi level and the McMillan formula systematically fails to yield the observed {\tc} variation. On the other hand, spin fluctuation theory pairing calculations allow us to consistently explain the {\tc} increase with doping. All alkali doped materials \ce{Ba$_{1-x}A_x$Ti2Sb2O} ($A$=Na, K, Rb) are described by a sign-changing $s$-wave order parameter. Susceptibilities also reveal that the physics of the materials is controlled by a single Ti $3d$ orbital.
\end{abstract}

\maketitle

{\it Introduction.-} The first layered titanium oxypnictides {\naas} and {\nasb} were synthesized three decades ago~\cite{Adam1990} and discussed in terms of spin density wave (SDW) or charge density wave (CDW) behavior~\cite{Axtell1997,Pickett1998}. Nine years ago, superconductivity was discovered in {\basb}~\cite{Yajima2012} and {\babi}~\cite{Yajima2013}. By analysis of the {\assb} solid solutions it quickly became apparent that superconductivity is favored by a suppression of the CDW/SDW phase~\cite{Yajima2013a,Zhai2013}. An extensive discussion of the phase transition observed in resistivity and magnetic susceptibility~\cite{Axtell1997,Wang2010,Yajima2012,Doan2012} as well as thermoelectric power and Hall coefficient~\cite{Liu2009} concerns the question whether it should be characterized as a CDW transition or if it is in fact an SDW transition. Experimental evidence from nuclear magnetic resonance (NMR)~\cite{Kitagawa2013} and muon spin relaxation ($\mu$SR)~\cite{vonRohr2013,Nozaki2013} does not completely resolve the question.

Shortly after the discovery of superconductivity, it was realized that both charge doping on the barium site and isovalent doping on the pnictogen site provide opportunities to control the superconductivity of {\basb} in a significant range. Hole doping via alkali metals increases {\tc} from 1.2\,K to 5.5\,K in {\bana}~\cite{Doan2012,vonRohr2013}, to 6.1\,K in {\bak}~\cite{Pachmayr2014}, to 5.4\,K in {\barb}~\cite{vonRohr2014} and to 4.4\,K in {\bacs}~\cite{Wang2019}. The maximum {\tc} is reached near an alkali content of $x=0.2$ to $0.3$. Isovalent doping via Sb/Bi mixing yields an intriguing two-dome {\tc} evolution with a non-superconducting or low {\tc} phase in between~\cite{Yajima2013a,Ishii2018}. More recently, pressure has been demonstrated to be an effective control parameter for superconductivity~\cite{Wang2020}. 

The nature of superconductivity in the titanium oxypnictides has been discussed since its discovery~\cite{Lorenz2014}. Experimentally, an $s$-wave gap has been inferred from nuclear quadrupole resonance (NQR) measurements~\cite{Kitagawa2013}, and specific heat is partially consistent with BCS expectations~\cite{Gooch2013}. However, $\mu$SR measurements have been taken to indicate an unconventional pairing mechanism~\cite{Kamusella2014}. An NMR/NQR study points to significant differences in the superconductivity of {\basb} and {\babi}~\cite{Kitagawa2018}. 

Theoretically, calculations of electron-phonon coupling have shown that the small transition temperature of {\basb} can be explained by an electron-phonon mechanism~\cite{Subedi2013}. On the other hand, based on the Fermi surfaces a sign-changing $s$-wave state has been predicted within spin fluctuation theory~\cite{Singh2012}. While the presence of magnetism has not been fully established in the titanium oxypnictides, an extensive density functional theory (DFT) study and symmetry analysis of the nematicity and charge order in {\basb} have provided strong evidence that these materials cannot be understood without taking spin fluctuations into account~\cite{Zhang2017}. 

While the electronic structure of individual titanium oxypnictide metals~\cite{Yu2014,Kim2017} and superconductors~\cite{Singh2012,Wang2013,Yan2013} has been studied repeatedly, there is no theory for the evolution of properties with doping. Our study is intended to fill this gap. In this work, we show that we can consistently explain the evolution of the superconducting {\tc} with alkali doping using spin fluctuation theory and that the superconducting gap function has a sign-changing $s$-wave symmetry.

\begin{figure}[htb]
    \includegraphics[width=\columnwidth]{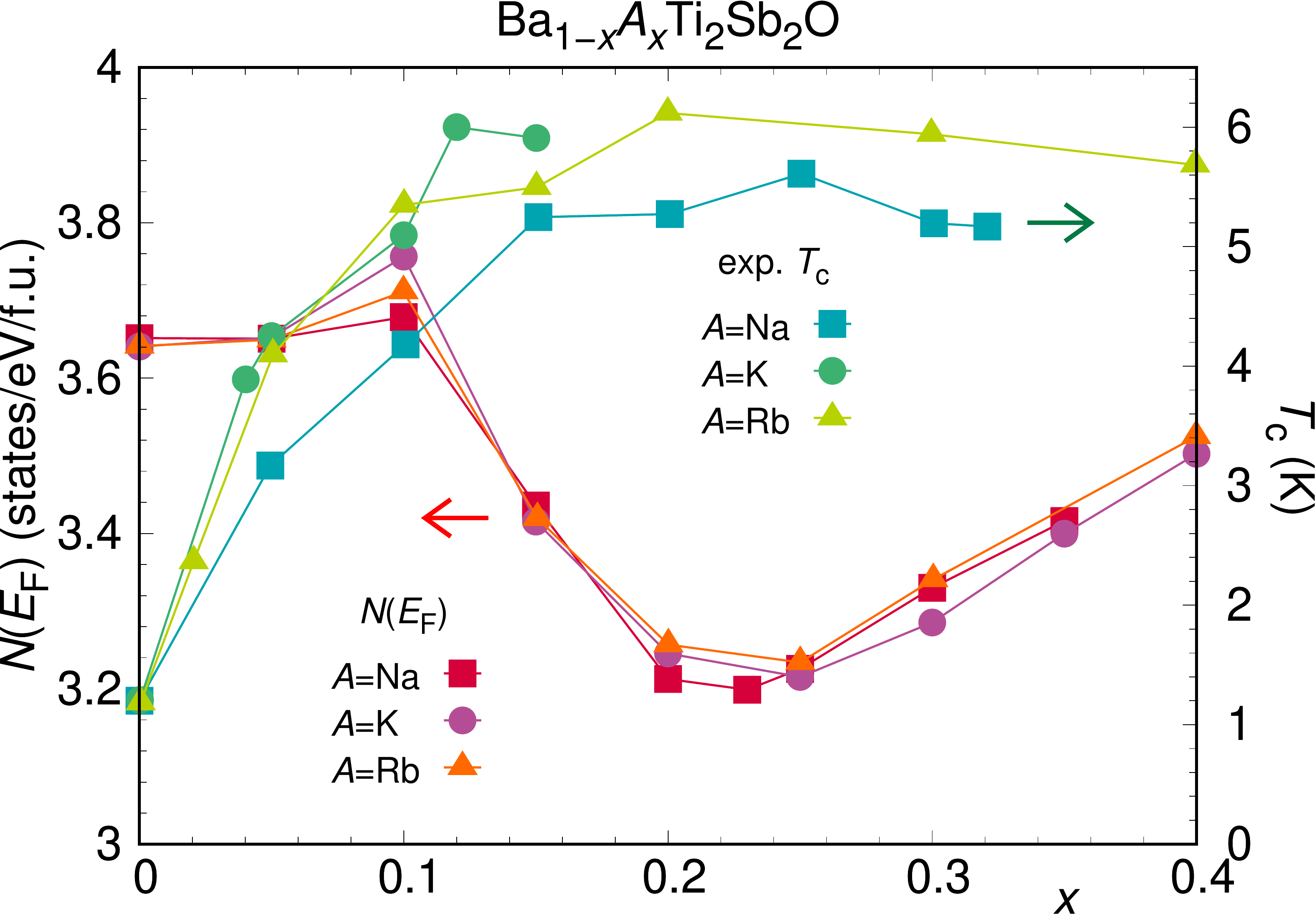}
    \caption{Density of states at the Fermi level $N(E_{\rm F})$ shown together with experimental superconducting transition temperatures {\tc} for doping with three different alkali ions. {\tc} data are from Ref.~\onlinecite{Doan2012} and \onlinecite{Gooch2013} for Na doping, Ref.~\onlinecite{Pachmayr2014} for K doping, Ref.~\onlinecite{vonRohr2014} for Rb doping.}
    \label{fig:nef}
\end{figure}

\begin{figure*}[htb]
    \includegraphics[width=\textwidth]{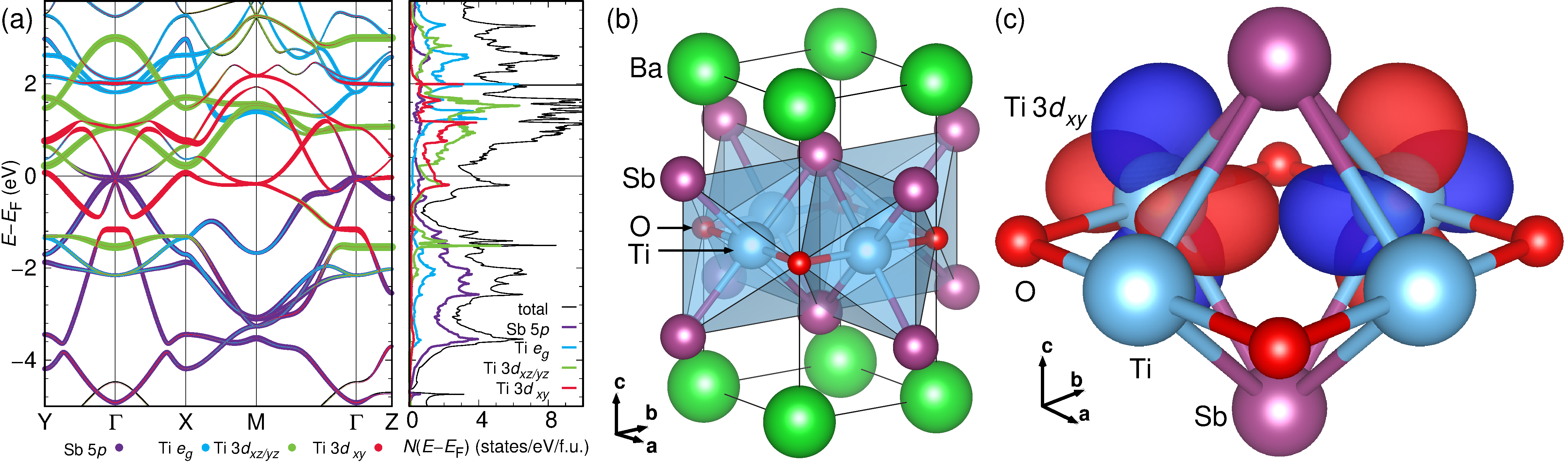}
    \caption{(a) GGA band structure and density of states of {\basb}. (b) Crystal structure of {\basb} with shaded \ce{TiO2Sb4} octahedra. (c) Ti $3d_{xy}$ Wannier functions within the \ce{Ti2Sb2O} layer of {\basb}.}
    \label{fig:wannier}
\end{figure*}

\begin{figure}[htb]
    \includegraphics[width=0.49\textwidth]{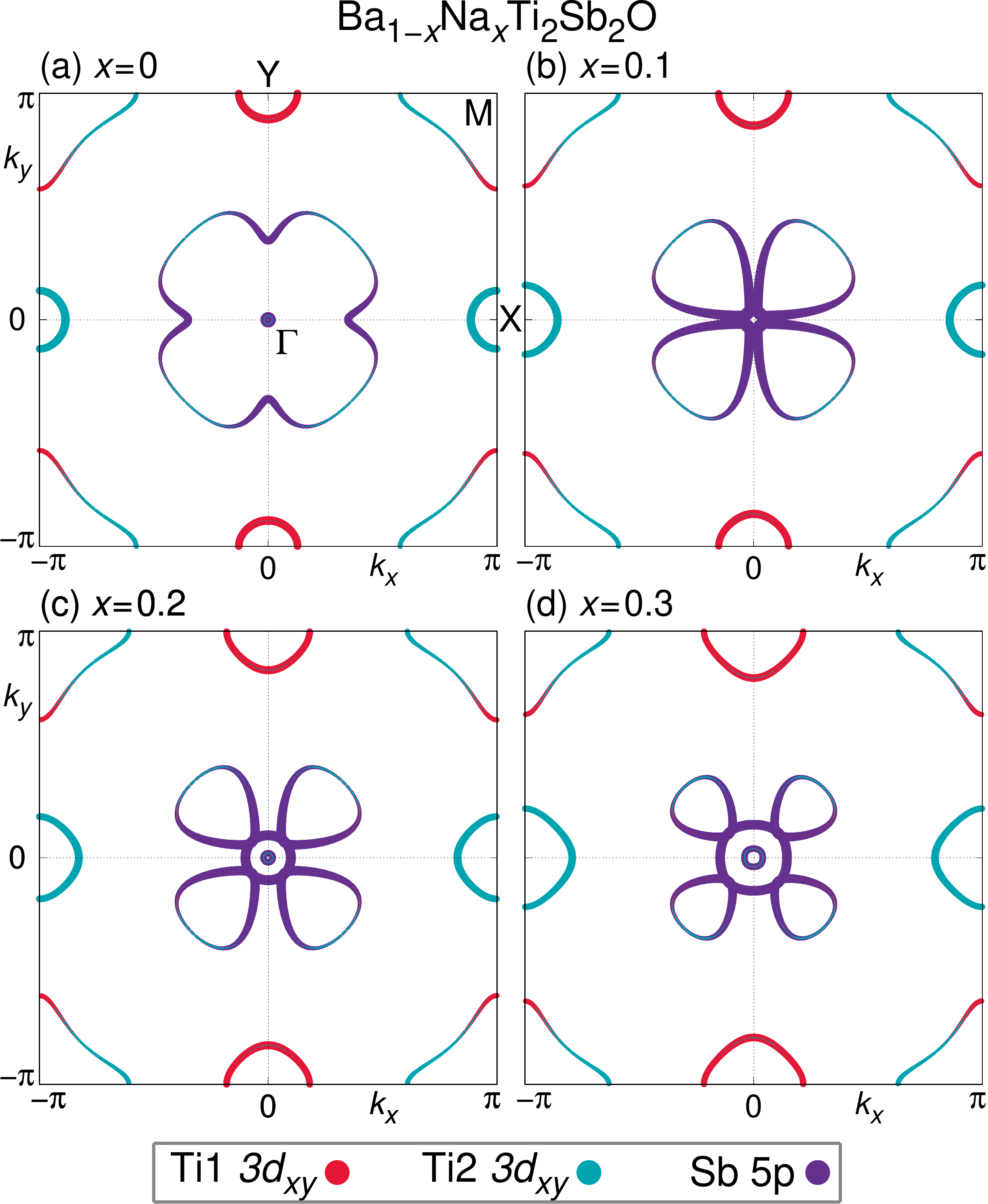}
    \caption{Fermi surfaces of {\bana} at $k_z=0$ as function of doping level $x$, calculated within GGA.}
    \label{fig:nafs}
\end{figure}

\begin{figure}[htb]
    \includegraphics[width=\columnwidth]{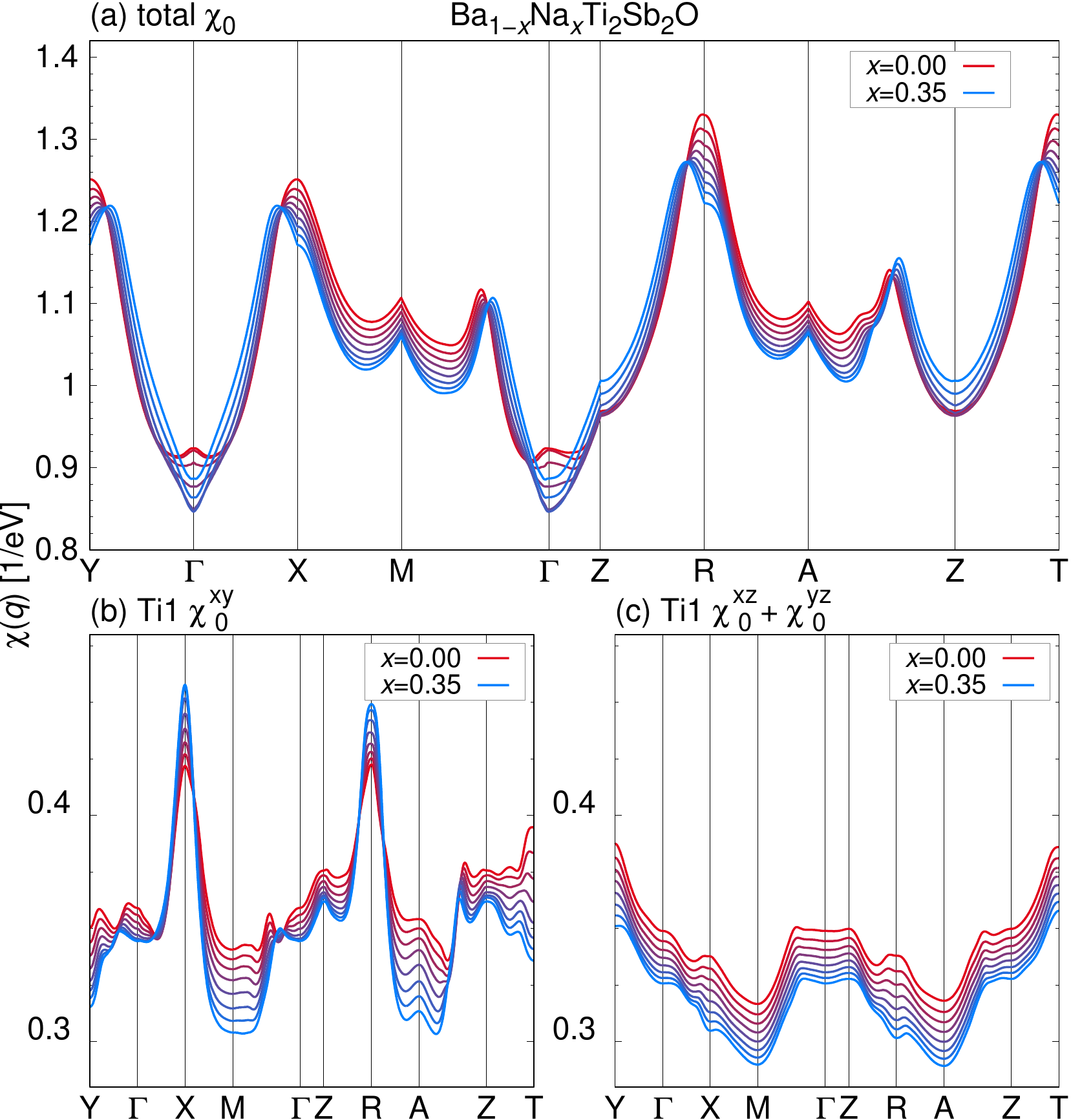}
    \caption{Non-interacting susceptibility of {\bana} for eight doping levels $x$. (a) total, (b) $3d_{xy}$ contribution from Ti1, (c) $3d_{xz}$ and $3d_{xz}$ contributions from Ti1. ${\bf q}=(\pi,0,0)$ is labeled as $X$, ${\bf q}=(0,\pi,0)$ as $Y$, ${\bf q}=(\pi,\pi,0)$ as $M$, ${\bf q}=(0,0,\pi)$ as $Z$, ${\bf q}=(0,\pi,\pi)$ as $R$, ${\bf q}=(\pi,\pi,\pi)$ as $A$ and ${\bf q}=(\pi,0,\pi)$ as $T$.}
    \label{fig:nasuscep}
\end{figure}

\begin{figure*}[htb]
    \includegraphics[width=0.99\textwidth]{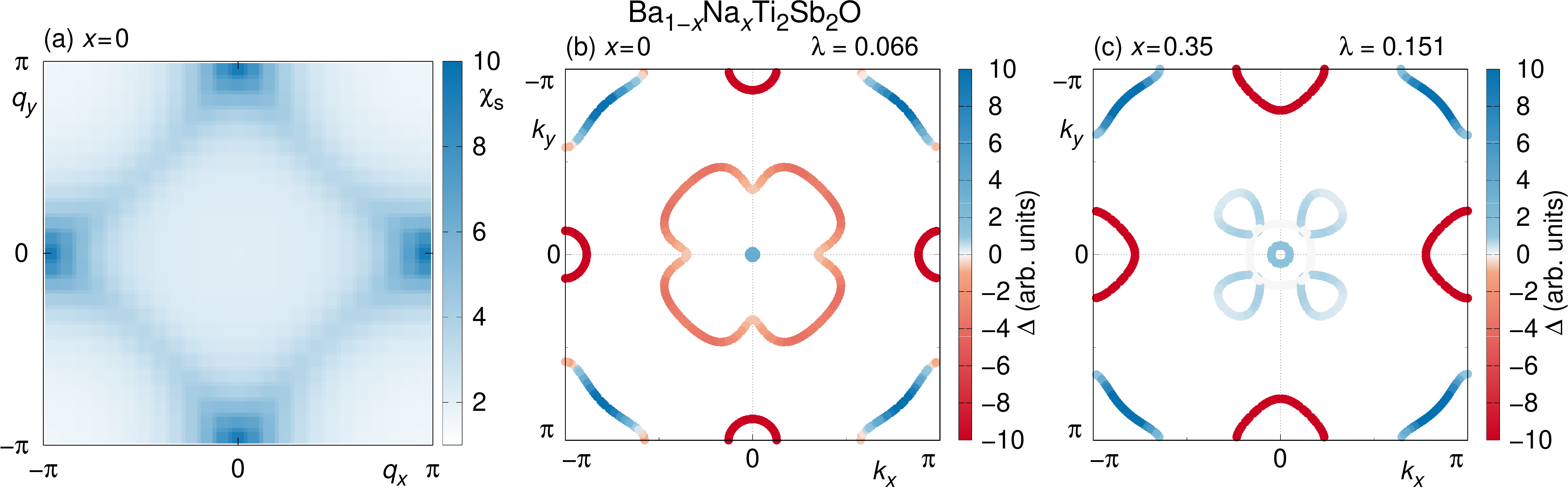}
    \caption{Two-dimensional susceptibility and gap functions of {\bana}. (a) Spin susceptibility $\chi_{\rm s}$ calculated within RPA for $x=0$, showing enhanced maxima at ${\bf q}=(0,\pi)$ and ${\bf q}=(\pi,0)$. (b) and (c) Eigen functions for the leading eigenvalue of the gap equation at zero doping and at maximal doping. The sign-changing $s$-wave persists at all doping levels. }
    \label{fig:pairing}
\end{figure*}

\begin{figure}[htb]
    \includegraphics[width=\columnwidth]{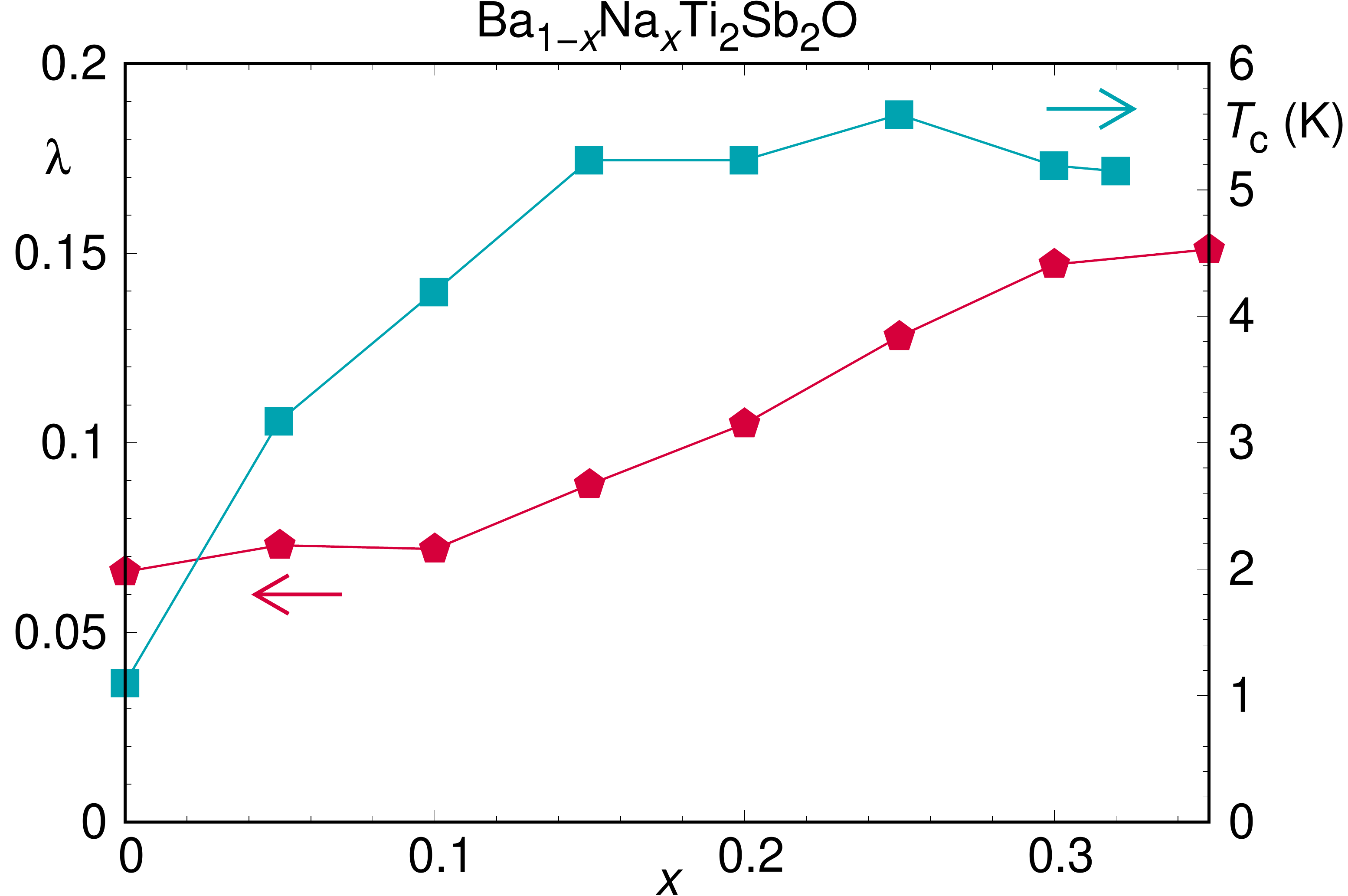}
    \caption{Leading eigenvalues $\lambda$ of the gap equation as a function of doping level $x$. {\tc} data are from Ref.~\onlinecite{Doan2012} and \onlinecite{Gooch2013}.}
    \label{fig:lambda}
\end{figure}

{\it Methods.-} We perform electronic structure calculations using the full potential local orbital (FPLO) basis set~\cite{Koepernik1999} and generalized gradient approximation (GGA) to the exchange and correlation potential~\cite{Perdew1996}. We use smooth interpolations of lattice parameters of the $P4/mmm$ space group  (see Appendix~\ref{app:structure}, Fig.~\ref{fig:structure}) 
and optimize the antimony positions within GGA. The charge doping is modeled via the virtual crystal approximation on the barium site. We use projective Wannier functions within FPLO~\cite{Eschrig2009} to obtain 26 band tight binding models including Ti $3d$ and $4s$, Ba $5d$, Sb $5p$ and O $2p$ orbital characters. The models follow DFT bands and Fermi surfaces to a high degree of accuracy (see Appendix~\ref{app:compare}, Fig.~\ref{fig:compare}). 
Based on the tight binding model, we calculate non-interacting susceptibilities $\chi^{pq}_{st}({\bf q})$. We apply the random phase approximation (RPA) and investigate the pairing instabilities within spin fluctuation theory by solving the gap equation on the Fermi surface~\cite{Guterding2015,Guterding2017,Shimizu2018,Shimizu2020} (for details, see Appendix~\ref{app:spinfluct}). Interaction parameters of $U=2$~eV (intra-orbital Coulomb repulsion), $U'=1$~eV (inter-orbital Coulomb repulsion), $J=0.5$~eV (Hund's rule coupling) and $J'=0.5$~eV (pair hopping), applied to Ti $3d$ orbitals, were used for RPA and pairing calculations. Note that a limitation of RPA based fluctuation theory is the restriction to zero energy, and application of the fluctuation exchange approximation (FLEX) which takes the energy dependence into account is an interesting future extension of our study.

{\it Results.-} The titanium oxypnictide superconductors have been treated as simple Bardeen, Cooper, Shrieffer (BCS) type superconductors in various experimental~\cite{Yajima2012,Kitagawa2013,vonRohr2014,Gooch2013,Hosono2015} and theoretical~\cite{Subedi2013} studies. As a straight-forward attempt to understand the {\tc} tendencies, we extract the density of states at the Fermi level {\nef} as a function of doping and try to apply the BCS formula $T_{\rm c}=1.134 \, T_{\rm D} \exp\big(-\frac{1}{VN(E_{\rm F})}\big)$ (with Debye temperature $T_{\rm D}$ and electron-phonon coupling potential $V$). Assuming constant $T_{\rm D}$ and $V$, this formula and its more sophisticated variants yield {\tc} trends that essentially follow {\nef}. Note that our use of the virtual crystal approximation is justified by good comparison of our Fermi surfaces to angle resolved photoemission (see Appendix~\ref{app:elstruct}, Fig.~\ref{fig:na0p05fs}). Nevertheless, conducting a similar study based on a more elaborate approach for treating alkali doping and Ba/alkali site disorder like the coherent potential approximation will be an interesting future endeavor.

Unfortunately, as Fig.~\ref{fig:nef} shows, there is very little similarity between {\nef} and {\tc} evolution with doping. In the case of alkali doping of {\basb}, {\tc} quickly increases from $T_{\rm c}=1.2$\,K to a maximum that is reached between doping levels of $x=0.1$ to 0.3. Meanwhile, {\nef} remains constant until $x=0.1$ before going through a minimum at $x=0.23$ (Fig.~\ref{fig:nef}). 
Based on this analysis, superconductivity in these materials could only be explained by an electron-phonon mechanism if the strength of electron-phonon coupling was extremely doping-dependent, so that it counteracts the unhelpful trends in {\nef}. However, this seems very far-fetched  because in the small doping range considered, neither $T_{\rm D}$ nor $V$ are expected to vary strongly. Therefore, we now turn to the possibility that the detailed evolution of the Fermi surface nesting provides doping dependencies strong enough to explain the evolution of {\tc} within spin fluctuation pairing theory. So far, the charge density wave state has been studied with spin fluctuation theory including Aslamazov-Larkin vertex corrections~\cite{Nakaoka2016} or within dynamical mean field theory~\cite{Song2018} but only a limited study of superconductivity exists~\cite{Wang2013}.

First, we identify the most relevant orbitals at the Fermi level. In Fig.~\ref{fig:wannier}\,(a), we show the band structure and density of states of {\basb} with Ti $3d$ and Sb $5p$ orbital character highlighted. The system is in general quite strongly hybridized and many orbitals contribute to the states close to the Fermi level. Taking a closer look, we find that the most relevant Ti $3d$ orbital for the low energy physics is $3d_{xy}$ followed by $3d_{xz,yz}$. To visualize the Ti $3d_{xy}$ orbital, we choose a local coordinate system for Ti where the $z$-axis points along the Ti-O bond and $x$- and $y$-axes point along Ti-Sb bonds. This is the natural local system to choose within the \ce{TiO2Sb4} octahedron~\cite{Kim2017} (Fig.~\ref{fig:wannier}\,(b)), since it makes $3d_{xz}$ and $3d_{yz}$ degenerate. Fig.~\ref{fig:wannier}\,(c) shows the $3d_{xy}$ Wannier functions at both titanium sites (Ti1 and Ti2) based on this coordinate choice.

We now analyze the Fermi surface evolution with alkali doping (Fig.~\ref{fig:nafs}). Only Ti $3d_{xy}$ and Sb $5p$ weights are highlighted  (see Appendix~\ref{app:elstruct}, Fig.~\ref{fig:fsorb} for the other $3d$ weights). Note that focusing on Ti and Sb is justified because relative contributions to the density of states at the Fermi level $N(E_{\rm F})$ are 74\%, 20\%, 4\% and 1\% for Ti, Sb, Ba and O, respectively. The {\basb} Fermi surface is in excellent agreement with angle resolved photoemission (ARPES) experiments~\cite{Davies2016,Huang2020} (see Appendix~\ref{app:elstruct}, Fig.~\ref{fig:na0p05fs}). 
We see that the hole Fermi surfaces at $X$ and $Y$ grow with doping while the electron Fermi surface at $M$ shrinks slightly. The Fermi surface at $\Gamma$, which is dominated by Sb $5p$, shows a rather complicated reconstruction as a function of doping. This can be understood by tracing which orbital fillings are depleted by the holes introduced as function of alkali doping level $x$. In fact, the majority of doped holes are in Sb orbitals while Ti $3d$ orbitals are nearly unaffected. The changes seen in Fermi surfaces with Ti $3d$ character (Fig.~\ref{fig:nafs}) are due to stronger Sb-Ti bonding upon hole doping rather than due to a Fermi level shift.

In order to measure the relative importance of the Fermi surface changes, we turn to non-interacting susceptibilities calculated with the 26 band tight binding models on $50\times 50\times 50$ integration meshes. Fig.~\ref{fig:nasuscep} shows that the total susceptibility $\chi_0$ is clearly peaked at ${\bf q}=(\pi,0,0)$ (labeled $X$) and ${\bf q}=(0,\pi,0)$ (labeled $Y$). Previously, this has been noted based on the Lindhard function calculated without matrix elements~\cite{Singh2012,Wang2013}. With alkali doping, the peaks at $X$ and $Y$ decrease, and they also move away from the high symmetry point towards $\Gamma$. At the same time, $\chi_0$ at ${\bf q}=(0,0,0)$ decreases. Interestingly, the doping trends of the total $\chi_0$ and $\chi_0^{xy}$ differ: While the ratio between $X$ and $\Gamma$ values of $\chi_0$ hardly changes with doping, this ratio sharply increases for $\chi_0^{xy}$ due to increases at $X$ combined with decreases at $\Gamma$. This improved nesting is shown in Fig.~\ref{fig:nasuscep}\,(b) for Ti1, but equally applies to Ti2 where the $Y$ to $\Gamma$ ratio increases sharply. Furthermore, even though the Fermi surface shows substantial $k_z$ dispersion, we can find the improved nesting also in the $R$ to $Z$ ratio. Meanwhile, the other orbitals, which are of some significance at the Fermi level ($3d_{xz}$ and $3d_{yz}$), have a comparatively featureless susceptibility (Fig.~\ref{fig:nasuscep}\,(c)) which uniformly decreases with doping. It is justified to focus on Ti $3d$ susceptibilities here because the Sb susceptibilities are small, almost flat with respect to {\bf q} and they grow even more featureless with doping (see Fig.~\ref{fig:chi0Sb}). Note that calculations without matrix elements, i.e. solely based on the Lindhard function, do not contain the orbital-resolved information we just discussed. Since alkali doping seems to lead to an overall decrease in susceptibility, but strongly enhances the susceptibility of the Ti $3d_{xy}$ orbitals, we can expect them to be the main actor in {\tc} changes with doping. 

Since the similarity of susceptibilities along the $k_z=0$ and $k_z=\pi$ paths in Fig.~\ref{fig:nasuscep}\,(a) indicates a high degree of two-dimensionality, we now focus on the $k_z=0$ cuts of susceptibility and pairing in Fig.~\ref{fig:pairing}. It is clear that features of the non-interacting susceptibility $\chi_0$ (see Appendix~\ref{app:suscept}, Fig.~\ref{fig:chi0}), especially peaks at ${\bf q}=(\pi,0)$ and ${\bf q}=(0,\pi)$, are enhanced in the interacting susceptibility obtained by random phase approximation (Fig.~\ref{fig:pairing}\,(a)), reminiscent of single orbital system behaviour. These instabilities would now favor stripe-type magnetism which, however, has not been observed for {\basb}~\cite{Nozaki2013,vonRohr2013,Kitagawa2013}. %They are also consistent with a small orthorhombic lattice distortion observed in neutron diffraction~\cite{Frandsen2014}. 

Here, the $(\pi,0)$ and $(0,\pi)$ instabilities favor a sign-changing $s$-wave superconducting order parameter: The gap functions corresponding to the leading eigenvalue $\lambda$, obtained using spin fluctuation theory, are shown in Fig.~\ref{fig:pairing}\,(b) and (c) for for two different doping levels. Subleading $d_{xy}$- and $d_{x^2-y^2}$-type solutions have far smaller eigenvalues and are, therefore, irrelevant in {\bana}. The eigenvalue $\lambda$ increases as a function of alkali doping (Fig.~\ref{fig:lambda}), and follows the doping trend of the maxima in the susceptibility (Fig.~\ref{fig:nasuscep}\,(a)). Thus, the increase of {\tc} with alkali doping (Fig.~\ref{fig:nef}) is clearly explained by the susceptibility trends rather than the density of states at the Fermi level. By performing the pairing calculations on 3D Fermi surfaces, we have verified that the sign-changing $s$-wave is indeed the dominating solution for all alkali doped materials \ce{Ba$_{1-x}A_x$Ti2Sb2O}  ($A$=Na, K, Rb).

At low alkali doping levels we have found an order parameter, which contains nodes on the Fermi surface sheets around $M$ and a relatively large gap on the central Fermi surface sheets around $\Gamma$. With increasing doping, the Fermi surface sheets around $M$ become nodeless, but the reconstructed sheets around $\Gamma$, which are almost exclusively of Sb $5p$ character, are hardly gapped at all. Those strongly non-uniform order parameters need to be taken into account when interpreting thermodynamic and other experiments trying to determine the symmetry of the superconducting state in titanium oxypnictides.

{\it Conclusions.-} We have investigated the electronic and superconducting properties of {\bana} using density functional and spin fluctuation theory. We modeled the crystal structure evolution using an interpolation of experimental lattice parameters and a DFT predicted antimony position. The density of states at the Fermi level {\nef} shows a trend which is in sharp contrast to the evolution of the superconducting {\tc}, indicating that transition temperatures may not be accounted for by an electron-phonon mechanism.

Although the band structure and density of states show that constituents of \ce{Ba$_{1-x}A_x$Ti2Sb2O} ($A$=Na, K, Rb) are strongly hybridized and many orbitals lie close to the Fermi level, we have found that the susceptibility is completely dominated by the Ti $3d_{xy}$ orbitals.

Proceeding on the assumption of a magnetic pairing mechanism, which has been suggested by an investigation into the nematicity of {\basb} in Ref.~\onlinecite{Zhang2017}, we find that we can satisfactorily explain the {\tc} trend with a spin fluctuation pairing mechanism. We find that a sign-changing $s$-wave order parameter with non-uniform gap size on the various Fermi surface sheets clearly dominates in {\bana} at all doping levels. Explaining the nontrivial transition temperature trends of titanium based superconductors with isoelectronic doping and pressure are interesting future fields of study. Methodologically, it may be important to consider also the energy dependence within the fluctuation exchange approximation (FLEX).

\appendix

\begin{figure}[htb]
    \includegraphics[width=\columnwidth]{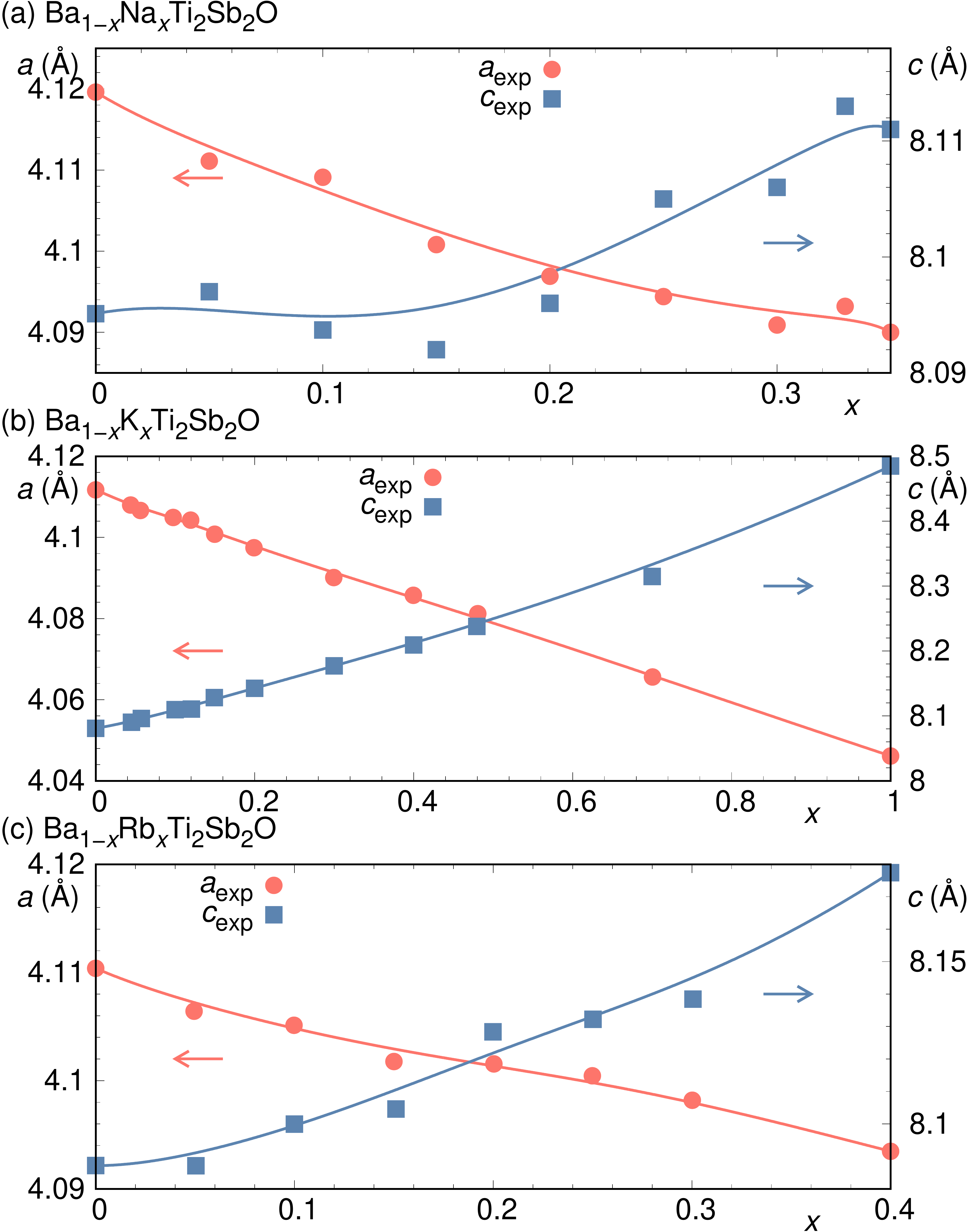}
    \caption{Interpolation of experimental lattice constants.}
    \label{fig:structure}
\end{figure}

\begin{figure}[htb]
    \includegraphics[width=\columnwidth]{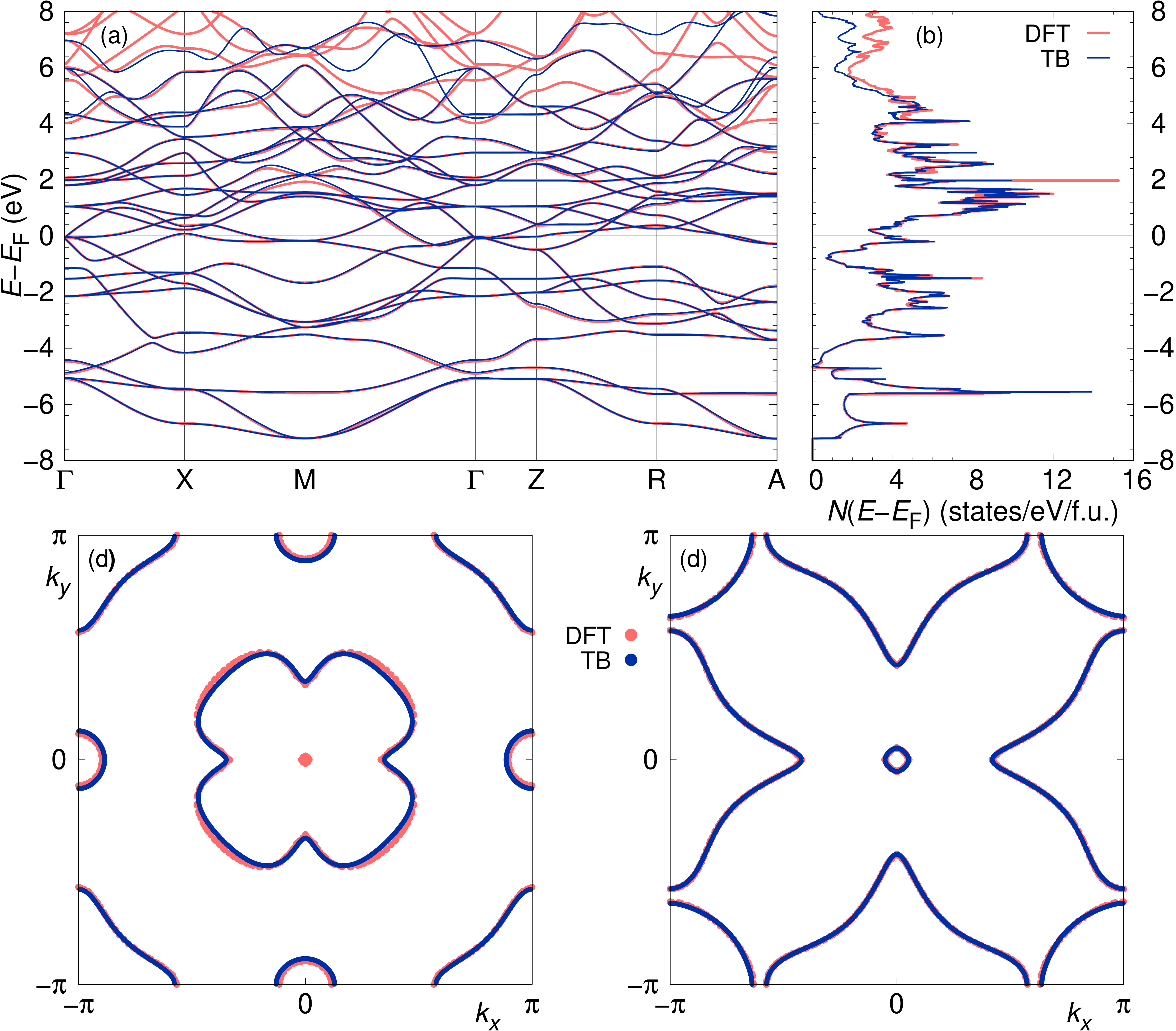}
    \caption{Comparison between density functional theory and tight binding model for {\basb}. (a) Band structure, (b) density of states, (c) Fermi surface at $k_{z}=0$ and (d) Fermi surface at $k_{z}=\pi$. The agreement is excellent.}
    \label{fig:compare}
\end{figure}

\begin{figure}[htb]
    \includegraphics[width=\columnwidth]{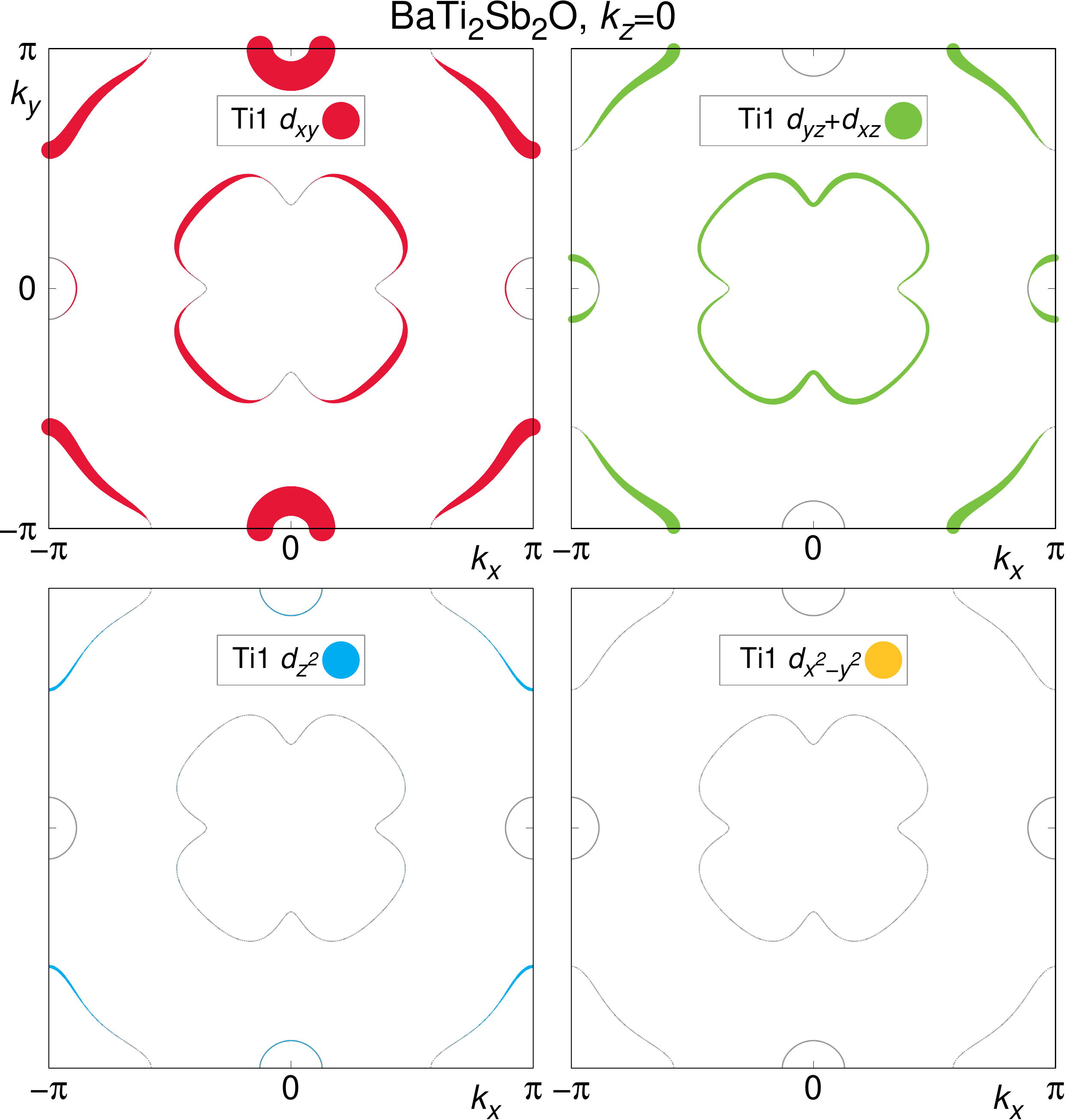}
    \caption{2D Fermi surface of {\basb} with Ti $3d$ orbital character. All weights are shown with the same scale. $3d_{xy}$ clearly dominates, followed in importance by $3d_{yz}$/$3d_{xz}$. $3d_{z^2}$ is very faint, and $3d_{x^2-y^2}$ character is negligible. Weights of the second Ti site are 90 degree rotated with respect to the first so that the sum has the $C_4$ symmetry of the space group.}
    \label{fig:fsorb}
\end{figure}

\begin{figure}[htb]
    \includegraphics[width=0.7\columnwidth]{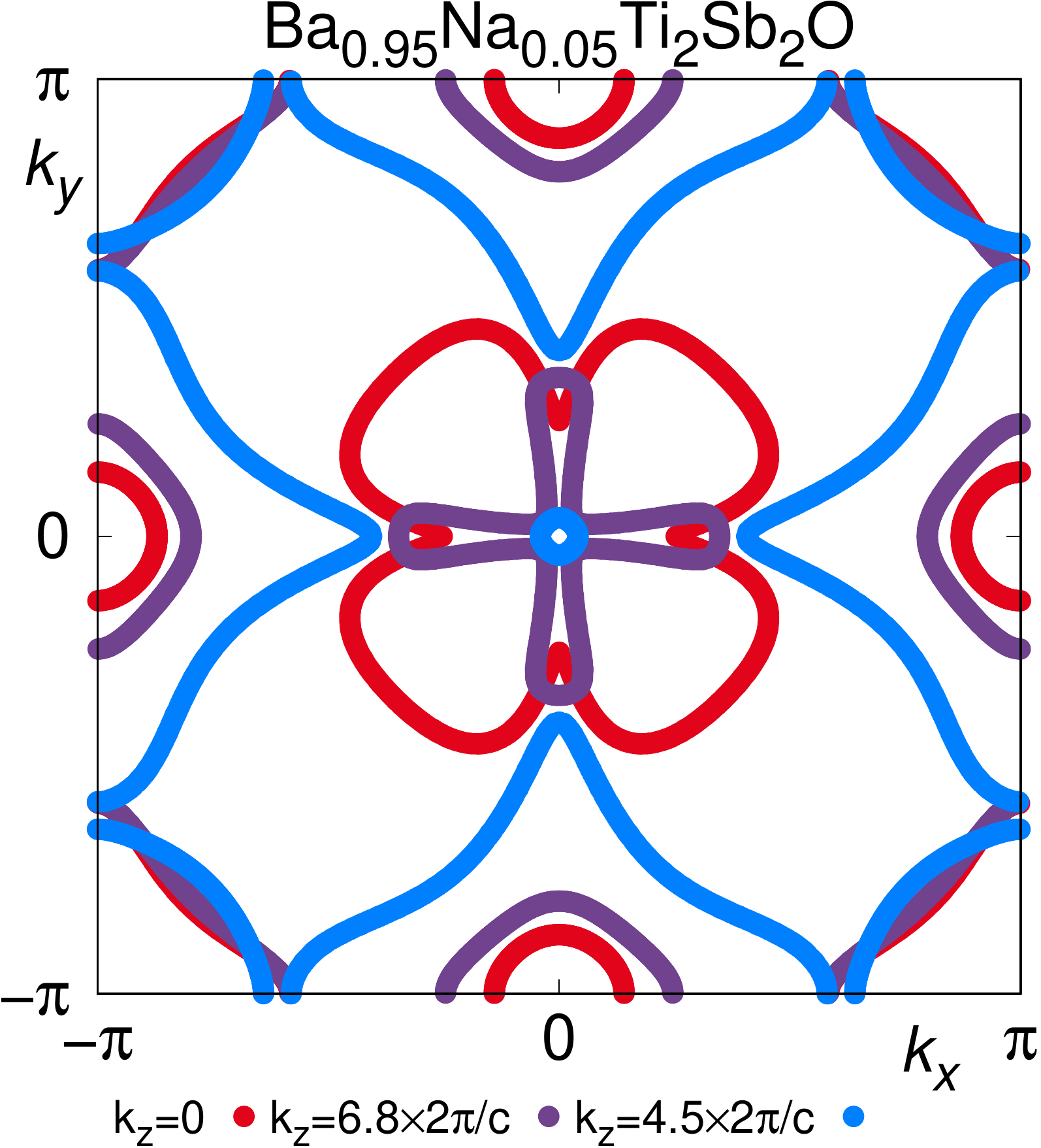}
    \caption{Fermi surface of \ce{Ba$_{0.95}$Na$_{0.05}$Ti2Sb2O} at $k_z=0$, $k_z=0.5\pi$ and $k_z=\pi$, calculated within GGA. }
    \label{fig:na0p05fs}
\end{figure}

\begin{figure*}[htb]
    \includegraphics[width=0.8\textwidth]{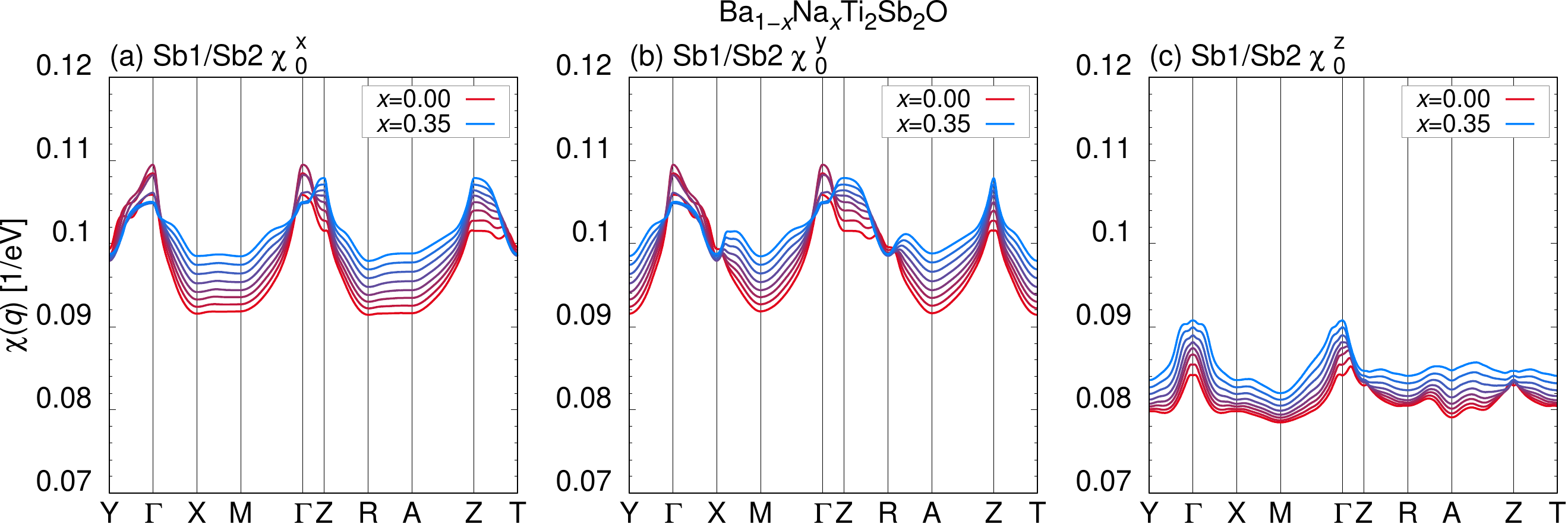}
    \caption{Sb contribution to the non-interacting susceptibility $\chi_0$ for {\bana} along a {\bf q} path. See Fig.~\ref{fig:nasuscep} for the meaning of the path labels.}
    \label{fig:chi0Sb}
\end{figure*}

\begin{figure}[htb]
    \includegraphics[width=\columnwidth]{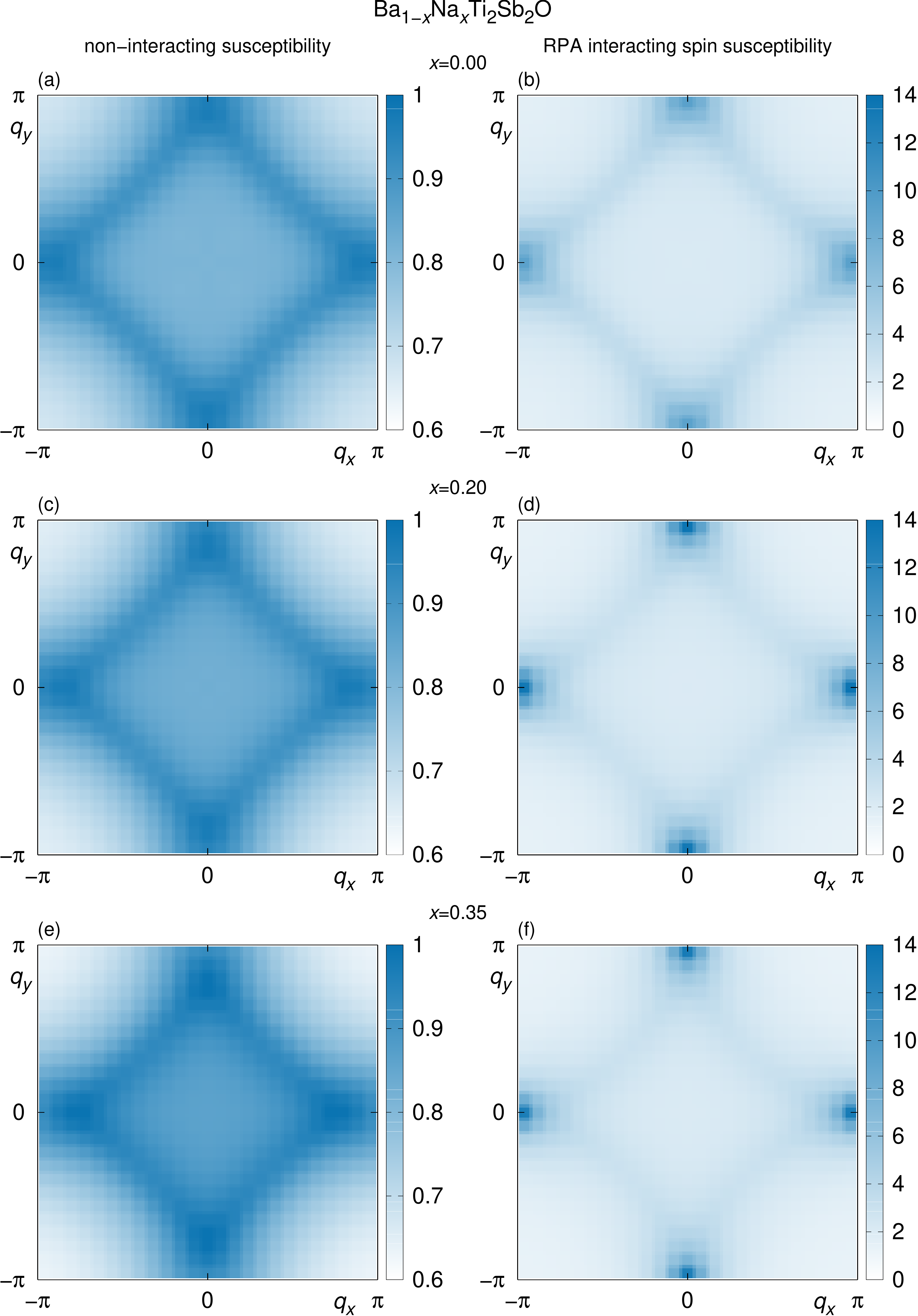}
    \caption{Non-interacting susceptibility $\chi_0$ and RPA interacting spin susceptibility $\chi_s$ for {\bana}. }
    \label{fig:chi0}
\end{figure}

\section{Crystal structures}\label{app:structure}

We use experimental lattice parameters for {\bana} from Ref.~\onlinecite{Doan2012}, for {\bak} from Ref.~\onlinecite{Pachmayr2014} and for {\barb} from Ref.~\onlinecite{vonRohr2014}. They are shown as symbols in Fig.~\ref{fig:structure}. We smoothly interpolate the lattice parameters in order to sample the doped crystal structures at regular intervals. Note that in the case of the $c$ lattice parameter of {\bana} where the experimental lattice constants are rather nonmonotonous, the overall scale of the variation is small; as there is little justification for a dramatic doping dependence, we expect that the slight smoothing due to the interpolation is reducing error rather than loosing detail. The antimony positions are the only free positions in the $P4/mmm$ crystal structures, and we obtain them by relaxation using FPLO basis~\cite{Koepernik1999} and GGA exchange correlation functional. Note that for the experimentally known {\basb} structure the deviation in Ti-Sb distance and Sb-Ti-Sb angle is only 0.2\% for the relaxed structure; this gives us confidence that the relaxation is reliable also for the doping series for which no experimental Sb position is available, in contrast to the well known difficulties of DFT structure prediction for iron based superconductors~\cite{Mazin2008}. We model the alkali doping $x$ by using the virtual crystal approximation for Ba, using a nuclear charge between $Z=55$ and $56$. 

\section{Tight binding model}\label{app:compare}

We use projective Wannier functions within FPLO~\cite{Eschrig2009} to construct faithful tight binding models $t_{ij}^{sp}$ of {\basb} and the alkali doping series:
\begin{equation}
	 H_{0} = -\sum_{i,j} t_{ij}^{sp} c^{\dag}_{is\sigma} c_{jp\sigma}
  \label{eq:HTB}
\end{equation}
where the $t_{ij}$ are transfer integrals between sites $i$ and
$j$, $s$ and $p$ are orbital indices, and $\sigma$ is the
spin. Fig.~\ref{fig:compare} shows the quality of fit for band structure, density of states and Fermi surface of {\basb}; the agreement is nearly perfect. To achieve this, we need to include 26 orbitals: Ten Ti $3d$ orbitals, two Ti $4s$ orbitals, six Sb $5p$ orbitals, five Ba $5d$ orbitals and three O $2p$ orbitals.

\section{Electronic structure}\label{app:elstruct}

Figure~\ref{fig:fsorb} shows the weight of all $3d$ orbitals of Ti1 for {\basb}. The Ti2 $3d$ orbitals have weights which are 90 degree rotated with respect to Ti1 (not shown). The dominating orbital is Ti $3d_{xy}$, and $3d_{xz}$, $3d_{yz}$ orbitals have some weight at the Fermi level as well. $3d_{z^2}$ and $3d_{x^2-y^2}$ orbitals contributions are negligibly small. 

Figure~\ref{fig:na0p05fs} shows the Fermi surface of {\bana} at $x=0.05$ and compares favorably with the angle resolved photoemission (ARPES) experiment of Ref.~\cite{Davies2016}.

\section{Spin fluctuation formalism}\label{app:spinfluct}

We consider the multiorbital Hubbard Hamiltonian~\cite{Graser2009}
\begin{equation}
  \begin{split}
	  H =& H_0 
	  	+ U \sum_{i,l} n_{il\uparrow} n_{il\downarrow}	+ \frac{U'}{2} \sum_{i,s,p{\neq}s} n_{is} n_{ip}\\
	  &
		- \frac{J}{2} \sum_{i,s,p{\neq}s} {\bm S}_{is} \cdot
		 {\bm S}_{ip} + \frac{J'}{2} \sum_{i,s,p{\neq}s, \sigma}
		 c^{\dag}_{is\sigma} c^{\dag}_{is\bar{\sigma}}
		 c_{ip\bar{\sigma}} c_{ip\sigma}
  \end{split}
  \label{eq:hamiltonian_parameters}
\end{equation}
with Fermionic creation (annihilation) operators $c^{\dag}_{is\sigma}$ $(c_{is\sigma})$, spin operator ${\bm S}_{is}$, density operator $n_{is\sigma}=c^{\dag}_{is\sigma} c_{is\sigma}$ and interaction parameters $U$, $U'$, $J$, $J'$ (intraorbital Coulomb repulsion, interorbital Coulomb
repulsion, Hund's rule coupling, pair-hopping term). The tight binding part $H_0$ is given by Eq.~\eqref{eq:HTB}. Diagonalization of $H_0$ provides band energies $E_l(\bm k)$ and matrix elements $a^{s}_{m}$ and allows calculation of the static noninteracting susceptibility
\begin{equation}
  \begin{split}
	 \chi^{pq}_{st}({\bm q}) =
	 &-\sum_{{\bm k}, l, m} a^{p*}_{l}({\bm k}) a^{t}_{l}({\bm k})
	 	a^{s*}_{m}({\bm k}+{\bm q}) a^{q}_{m}({\bm k}+{\bm q}) \\
	 &\times \frac{n_{F}(E_{l}({\bm k}))-n_{F}(E_{m}({\bm k}+{\bm q}))}
	 {E_{l}({\bm k})-E_{m}({\bm k}+{\bm q})}
  \end{split}
  \label{eq:noninteracting_suscep}
\end{equation}
$n_F(E)$ is the Fermi distribution function. The observable static susceptibility can be calculated as
\begin{equation}
\chi_0({\bm q})=\frac{1}{2} \sum_{ab}\chi^{bb}_{aa}({\bm q})
  \label{eq:chi0}
\end{equation}
Applying the random phase approximation~(RPA),
charge and spin susceptibilities are calculated from the
noninteracting susceptibility as
\begin{equation}
  \begin{split}
   \left[ (\chi_\text{c}^{RPA})_{st}^{pq} \right]^{-1}
	&= \left[ \chi_{st}^{pq} \right]^{-1}
	 + (U_\text{c})_{st}^{pq} \\
   \left[ (\chi_\text{s}^{RPA})_{st}^{pq} \right]^{-1}
	&= \left[ \chi_{st}^{pq} \right]^{-1} - (U_\text{s})_{st}^{pq}
  \end{split}
  \label{eq:RPAsuscep}
\end{equation}
where nonzero components of the multi-orbital Hubbard model interaction tensors are~\cite{Graser2009}
\begin{align}
	(U_\text{c})^{aa}_{aa} &= U
	&(U_\text{c})^{aa}_{bb} &= 2U', \notag \\
	(U_\text{c})^{ab}_{ab} &= \frac{3}{4}J-U'
	&(U_\text{c})^{ba}_{ab} &= J' \notag \\
	(U_\text{s})^{aa}_{aa} &= U
	&(U_\text{s})^{aa}_{bb} &= \frac{1}{2}J \notag \\
	(U_\text{s})^{ab}_{ab} &= \frac{1}{4}J+U'
	&(U_\text{s})^{ba}_{ab} &= J', 
  \label{eq:interaction_tensors}
\end{align}
 Then, the superconducting pairing vertex in the singlet channel is
 \begin{equation}
  \begin{split}
	\Gamma_{st}^{pq} ({\bf k}, {\bf k}^\prime)  =& \Biggl[\frac{3}{2} U_s \, \chi_s^{RPA} ({\bf k}-{\bf k}^\prime) 
	\, U_s + \frac{1}{2} U_s \\
	& - \frac{1}{2} U_c \, \chi_c^{RPA} ({\bf k} -
	 {\bf k}^\prime) U_c + \frac{1}{2} U_c \Biggr] _{ps}^{tq}.
  \end{split}
 \end{equation}
 This vertex in orbital space is projected onto 
 band space using the eigenvectors of $H_0$,
  \begin{equation}
  \begin{split}
	\Gamma_{ij}& ({\bf k}, {\bf k}^\prime) \\
	&= \sum \limits_{s,t,p,q}
	a_{ i}^{t *} (- {\bf k}) a_{ i}^{s *} ({\bf k}) \mathrm{Re} \left[ 
	\Gamma_{st}^{pq} ({\bf k}, {\bf k}^\prime) \right] a_{ j}^{p}
	({\bf k}^\prime) a_{ j}^{q} (- {\bf k}^\prime).
  \end{split}
 \end{equation}
and the gap equation
 \begin{equation}
  \begin{split}
	 &- \sum_{j} \oint_{C_j} \frac{dk^{\prime}_{\parallel}}{2\pi}
	  \frac{1}{4\pi v_{F} ({\bf k}^\prime)} \left[ \Gamma_{ij}
	 ({\bf k}, {\bf k}^\prime) + \Gamma_{i j} ({\bf k}, - {\bf k}^\prime)
	 \right]  g_j ({\bf k}^\prime) \\
	  &\quad= \lambda_i g_i({\bf k})
  \label{eq:gapequation}
  \end{split}
 \end{equation}
 is solved for the pairing eigenvalue $\lambda_i$ and the gap function $g_{i}({\bf k})$.

\section{Susceptibility}\label{app:suscept}

The evolution of the Sb $5p_x$, $5p_y$ and $5p_z$ contributions to the non-interacting susceptibility of {\bana} with doping level $x$ is shown in Fig.~\ref{fig:chi0Sb}. The values of the susceptibilities compared to Ti $3d$ orbitals are small (see scale of Fig.~\ref{fig:nasuscep}), and their variation with {\bf q} value is tiny. With doping, the three susceptibilities grow even more featureless.

The non-interacting susceptibility $\chi_0$ and and RPA interacting spin susceptibility $\chi_s$ is shown in Fig.~\ref{fig:chi0}. This is based on a 3D calculation of $\chi_0$ on a $25\times 25\times 9$ grid using a $25\times 25\times 9$ integration grid. The $q_z=0$ cut shown in Fig.~\ref{fig:chi0} is extracted using a $100\times 100$ interpolation grid. The same interpolation grid was used for the interacting susceptibility $\chi_s$ shown in the main text, Fig. 5\,(a).

\acknowledgments
We acknowledge fruitful discussions with Yoshihiro Kubozono. Part of the
computations were carried out at the Supercomputer Center at
the Institute for Solid State Physics, the University of Tokyo.
This work was supported by MEXT Leading Initiative for
Excellent Young Researchers.

%{\it Not yet cited.-}
%review~\cite{Yajima2017}

%{\basb} volume collapse~\cite{Yamamoto2020}

%{\babi} SC~\cite{Yajima2014}

%{\bana}~\cite{Litvinchuk2013,Song2016,Gooch2013a}

%theory~\cite{Shi2019,Nakano2016,Nakano2017}

%\bibliography{Tioxypnictides}% Produces the bibliography via BibTeX.

\begin{thebibliography}{43}%
\makeatletter
\providecommand \@ifxundefined [1]{%
 \@ifx{#1\undefined}
}%
\providecommand \@ifnum [1]{%
 \ifnum #1\expandafter \@firstoftwo
 \else \expandafter \@secondoftwo
 \fi
}%
\providecommand \@ifx [1]{%
 \ifx #1\expandafter \@firstoftwo
 \else \expandafter \@secondoftwo
 \fi
}%
\providecommand \natexlab [1]{#1}%
\providecommand \enquote  [1]{``#1''}%
\providecommand \bibnamefont  [1]{#1}%
\providecommand \bibfnamefont [1]{#1}%
\providecommand \citenamefont [1]{#1}%
\providecommand \href@noop [0]{\@secondoftwo}%
\providecommand \href [0]{\begingroup \@sanitize@url \@href}%
\providecommand \@href[1]{\@@startlink{#1}\@@href}%
\providecommand \@@href[1]{\endgroup#1\@@endlink}%
\providecommand \@sanitize@url [0]{\catcode `\\12\catcode `\$12\catcode
  `\&12\catcode `\#12\catcode `\^12\catcode `\_12\catcode `\%12\relax}%
\providecommand \@@startlink[1]{}%
\providecommand \@@endlink[0]{}%
\providecommand \url  [0]{\begingroup\@sanitize@url \@url }%
\providecommand \@url [1]{\endgroup\@href {#1}{\urlprefix }}%
\providecommand \urlprefix  [0]{URL }%
\providecommand \Eprint [0]{\href }%
\providecommand \doibase [0]{https://doi.org/}%
\providecommand \selectlanguage [0]{\@gobble}%
\providecommand \bibinfo  [0]{\@secondoftwo}%
\providecommand \bibfield  [0]{\@secondoftwo}%
\providecommand \translation [1]{[#1]}%
\providecommand \BibitemOpen [0]{}%
\providecommand \bibitemStop [0]{}%
\providecommand \bibitemNoStop [0]{.\EOS\space}%
\providecommand \EOS [0]{\spacefactor3000\relax}%
\providecommand \BibitemShut  [1]{\csname bibitem#1\endcsname}%
\let\auto@bib@innerbib\@empty
%</preamble>
\bibitem [{\citenamefont {Adam}\ and\ \citenamefont
  {Schuster}(1990)}]{Adam1990}%
  \BibitemOpen
  \bibfield  {author} {\bibinfo {author} {\bibfnamefont {A.}~\bibnamefont
  {Adam}}\ and\ \bibinfo {author} {\bibfnamefont {H.-U.}\ \bibnamefont
  {Schuster}},\ }\bibfield  {title} {\bibinfo {title} {Darstellung und
  kristallstruktur der pnictidoxide \ce{Na2Ti2As2O} und \ce{Na2Ti2Sb2O}},\
  }\href@noop {} {\bibfield  {journal} {\bibinfo  {journal} {Z. Anorg. Allg.
  Chem.}\ }\textbf {\bibinfo {volume} {584}},\ \bibinfo {pages} {150} (\bibinfo
  {year} {1990})}\BibitemShut {NoStop}%
\bibitem [{\citenamefont {Axtell}\ \emph {et~al.}(1997)\citenamefont {Axtell},
  \citenamefont {Ozawa}, \citenamefont {Kauzlarich},\ and\ \citenamefont
  {Singh}}]{Axtell1997}%
  \BibitemOpen
  \bibfield  {author} {\bibinfo {author} {\bibfnamefont {E.}~\bibnamefont
  {Axtell}}, \bibinfo {author} {\bibfnamefont {T.}~\bibnamefont {Ozawa}},
  \bibinfo {author} {\bibfnamefont {S.~M.}\ \bibnamefont {Kauzlarich}},\ and\
  \bibinfo {author} {\bibfnamefont {R.~R.}\ \bibnamefont {Singh}},\ }\bibfield
  {title} {\bibinfo {title} {Phase transition and spin-gap behavior in a
  layered tetragonal pnictide oxide},\ }\href@noop {} {\bibfield  {journal}
  {\bibinfo  {journal} {J. Solid State Chem.}\ }\textbf {\bibinfo {volume}
  {134}},\ \bibinfo {pages} {423} (\bibinfo {year} {1997})}\BibitemShut
  {NoStop}%
\bibitem [{\citenamefont {Pickett}(1998)}]{Pickett1998}%
  \BibitemOpen
  \bibfield  {author} {\bibinfo {author} {\bibfnamefont {W.~E.}\ \bibnamefont
  {Pickett}},\ }\bibfield  {title} {\bibinfo {title} {Electronic instability in
  inverse-\ce{K2NiF4}-structure \ce{Na2Sb2Ti2O}},\ }\href@noop {} {\bibfield
  {journal} {\bibinfo  {journal} {Phys. Rev. B}\ }\textbf {\bibinfo {volume}
  {58}},\ \bibinfo {pages} {4335} (\bibinfo {year} {1998})}\BibitemShut
  {NoStop}%
\bibitem [{\citenamefont {Yajima}\ \emph {et~al.}(2012)\citenamefont {Yajima},
  \citenamefont {Nakano}, \citenamefont {Takeiri}, \citenamefont {Ono},
  \citenamefont {Hosokoshi}, \citenamefont {Matsushita}, \citenamefont
  {Hester}, \citenamefont {Kobayashi},\ and\ \citenamefont
  {Kageyama}}]{Yajima2012}%
  \BibitemOpen
  \bibfield  {author} {\bibinfo {author} {\bibfnamefont {T.}~\bibnamefont
  {Yajima}}, \bibinfo {author} {\bibfnamefont {K.}~\bibnamefont {Nakano}},
  \bibinfo {author} {\bibfnamefont {F.}~\bibnamefont {Takeiri}}, \bibinfo
  {author} {\bibfnamefont {T.}~\bibnamefont {Ono}}, \bibinfo {author}
  {\bibfnamefont {Y.}~\bibnamefont {Hosokoshi}}, \bibinfo {author}
  {\bibfnamefont {Y.}~\bibnamefont {Matsushita}}, \bibinfo {author}
  {\bibfnamefont {J.}~\bibnamefont {Hester}}, \bibinfo {author} {\bibfnamefont
  {Y.}~\bibnamefont {Kobayashi}},\ and\ \bibinfo {author} {\bibfnamefont
  {H.}~\bibnamefont {Kageyama}},\ }\bibfield  {title} {\bibinfo {title}
  {Superconductivity in \ce{BaTi2Sb2O} with a $d^1$ square lattice},\
  }\href@noop {} {\bibfield  {journal} {\bibinfo  {journal} {J. Phys. Soc.
  Jpn.}\ }\textbf {\bibinfo {volume} {81}},\ \bibinfo {pages} {103706}
  (\bibinfo {year} {2012})}\BibitemShut {NoStop}%
\bibitem [{\citenamefont {Yajima}\ \emph
  {et~al.}(2013{\natexlab{a}})\citenamefont {Yajima}, \citenamefont {Nakano},
  \citenamefont {Takeiri}, \citenamefont {Hester}, \citenamefont {Yamamoto},
  \citenamefont {Kobayashi}, \citenamefont {Tsuji}, \citenamefont {Kim},
  \citenamefont {Fujiwara},\ and\ \citenamefont {Kageyama}}]{Yajima2013}%
  \BibitemOpen
  \bibfield  {author} {\bibinfo {author} {\bibfnamefont {T.}~\bibnamefont
  {Yajima}}, \bibinfo {author} {\bibfnamefont {K.}~\bibnamefont {Nakano}},
  \bibinfo {author} {\bibfnamefont {F.}~\bibnamefont {Takeiri}}, \bibinfo
  {author} {\bibfnamefont {J.}~\bibnamefont {Hester}}, \bibinfo {author}
  {\bibfnamefont {T.}~\bibnamefont {Yamamoto}}, \bibinfo {author}
  {\bibfnamefont {Y.}~\bibnamefont {Kobayashi}}, \bibinfo {author}
  {\bibfnamefont {N.}~\bibnamefont {Tsuji}}, \bibinfo {author} {\bibfnamefont
  {J.}~\bibnamefont {Kim}}, \bibinfo {author} {\bibfnamefont {A.}~\bibnamefont
  {Fujiwara}},\ and\ \bibinfo {author} {\bibfnamefont {H.}~\bibnamefont
  {Kageyama}},\ }\bibfield  {title} {\bibinfo {title} {Synthesis and physical
  properties of the new oxybismuthides \ce{BaTi2Bi2O} and \ce{(SrF)2Ti2Bi2O}
  with a $d^1$ square net},\ }\href@noop {} {\bibfield  {journal} {\bibinfo
  {journal} {J. Phys. Soc. Jpn.}\ }\textbf {\bibinfo {volume} {82}},\ \bibinfo
  {pages} {013703} (\bibinfo {year} {2013}{\natexlab{a}})}\BibitemShut
  {NoStop}%
\bibitem [{\citenamefont {Yajima}\ \emph
  {et~al.}(2013{\natexlab{b}})\citenamefont {Yajima}, \citenamefont {Nakano},
  \citenamefont {Takeiri}, \citenamefont {Nozaki}, \citenamefont {Kobayashi},\
  and\ \citenamefont {Kageyama}}]{Yajima2013a}%
  \BibitemOpen
  \bibfield  {author} {\bibinfo {author} {\bibfnamefont {T.}~\bibnamefont
  {Yajima}}, \bibinfo {author} {\bibfnamefont {K.}~\bibnamefont {Nakano}},
  \bibinfo {author} {\bibfnamefont {F.}~\bibnamefont {Takeiri}}, \bibinfo
  {author} {\bibfnamefont {Y.}~\bibnamefont {Nozaki}}, \bibinfo {author}
  {\bibfnamefont {Y.}~\bibnamefont {Kobayashi}},\ and\ \bibinfo {author}
  {\bibfnamefont {H.}~\bibnamefont {Kageyama}},\ }\bibfield  {title} {\bibinfo
  {title} {Two superconducting phases in the isovalent solid solutions
  \ce{BaTi2$Pn$2O} ({$Pn$} = {As}, {Sb}, and {Bi})},\ }\href@noop {} {\bibfield
   {journal} {\bibinfo  {journal} {J. Phys. Soc. Jpn.}\ }\textbf {\bibinfo
  {volume} {82}},\ \bibinfo {pages} {033705} (\bibinfo {year}
  {2013}{\natexlab{b}})}\BibitemShut {NoStop}%
\bibitem [{\citenamefont {Zhai}\ \emph {et~al.}(2013)\citenamefont {Zhai},
  \citenamefont {Jiao}, \citenamefont {Sun}, \citenamefont {Bao}, \citenamefont
  {Jiang}, \citenamefont {Yang}, \citenamefont {Tang}, \citenamefont {Tao},
  \citenamefont {Xu}, \citenamefont {Li}, \citenamefont {Cao}, \citenamefont
  {Dai}, \citenamefont {Xu},\ and\ \citenamefont {Cao}}]{Zhai2013}%
  \BibitemOpen
  \bibfield  {author} {\bibinfo {author} {\bibfnamefont {H.-F.}\ \bibnamefont
  {Zhai}}, \bibinfo {author} {\bibfnamefont {W.-H.}\ \bibnamefont {Jiao}},
  \bibinfo {author} {\bibfnamefont {Y.-L.}\ \bibnamefont {Sun}}, \bibinfo
  {author} {\bibfnamefont {J.-K.}\ \bibnamefont {Bao}}, \bibinfo {author}
  {\bibfnamefont {H.}~\bibnamefont {Jiang}}, \bibinfo {author} {\bibfnamefont
  {X.-J.}\ \bibnamefont {Yang}}, \bibinfo {author} {\bibfnamefont {Z.-T.}\
  \bibnamefont {Tang}}, \bibinfo {author} {\bibfnamefont {Q.}~\bibnamefont
  {Tao}}, \bibinfo {author} {\bibfnamefont {X.-F.}\ \bibnamefont {Xu}},
  \bibinfo {author} {\bibfnamefont {Y.-K.}\ \bibnamefont {Li}}, \bibinfo
  {author} {\bibfnamefont {C.}~\bibnamefont {Cao}}, \bibinfo {author}
  {\bibfnamefont {J.-H.}\ \bibnamefont {Dai}}, \bibinfo {author} {\bibfnamefont
  {Z.-A.}\ \bibnamefont {Xu}},\ and\ \bibinfo {author} {\bibfnamefont {G.-H.}\
  \bibnamefont {Cao}},\ }\bibfield  {title} {\bibinfo {title}
  {Superconductivity, charge- or spin-density wave, and metal-nonmetal
  transition in \ce{BaTi2(Sb$_{1-x}$Bi$_x$)2O}},\ }\href@noop {} {\bibfield
  {journal} {\bibinfo  {journal} {Phys. Rev. B}\ }\textbf {\bibinfo {volume}
  {87}},\ \bibinfo {pages} {100502(R)} (\bibinfo {year} {2013})}\BibitemShut
  {NoStop}%
\bibitem [{\citenamefont {Wang}\ \emph {et~al.}(2010)\citenamefont {Wang},
  \citenamefont {Yan}, \citenamefont {Ying}, \citenamefont {Li}, \citenamefont
  {Zhang}, \citenamefont {Xu},\ and\ \citenamefont {Chen}}]{Wang2010}%
  \BibitemOpen
  \bibfield  {author} {\bibinfo {author} {\bibfnamefont {X.~F.}\ \bibnamefont
  {Wang}}, \bibinfo {author} {\bibfnamefont {Y.~J.}\ \bibnamefont {Yan}},
  \bibinfo {author} {\bibfnamefont {J.~J.}\ \bibnamefont {Ying}}, \bibinfo
  {author} {\bibfnamefont {Q.~J.}\ \bibnamefont {Li}}, \bibinfo {author}
  {\bibfnamefont {M.}~\bibnamefont {Zhang}}, \bibinfo {author} {\bibfnamefont
  {N.}~\bibnamefont {Xu}},\ and\ \bibinfo {author} {\bibfnamefont {X.~H.}\
  \bibnamefont {Chen}},\ }\bibfield  {title} {\bibinfo {title} {Structure and
  physical properties for a new layered pnictide-oxide: \ce{BaTi2As2O}},\
  }\href@noop {} {\bibfield  {journal} {\bibinfo  {journal} {J. Phys.: Condens.
  Matter}\ }\textbf {\bibinfo {volume} {22}},\ \bibinfo {pages} {075702}
  (\bibinfo {year} {2010})}\BibitemShut {NoStop}%
\bibitem [{\citenamefont {Doan}\ \emph {et~al.}(2012)\citenamefont {Doan},
  \citenamefont {Gooch}, \citenamefont {Tang}, \citenamefont {Lorenz},
  \citenamefont {Möller}, \citenamefont {Tapp}, \citenamefont {Chu},\ and\
  \citenamefont {Guloy}}]{Doan2012}%
  \BibitemOpen
  \bibfield  {author} {\bibinfo {author} {\bibfnamefont {P.}~\bibnamefont
  {Doan}}, \bibinfo {author} {\bibfnamefont {M.}~\bibnamefont {Gooch}},
  \bibinfo {author} {\bibfnamefont {Z.}~\bibnamefont {Tang}}, \bibinfo {author}
  {\bibfnamefont {B.}~\bibnamefont {Lorenz}}, \bibinfo {author} {\bibfnamefont
  {A.}~\bibnamefont {Möller}}, \bibinfo {author} {\bibfnamefont
  {J.}~\bibnamefont {Tapp}}, \bibinfo {author} {\bibfnamefont {P.~C.~W.}\
  \bibnamefont {Chu}},\ and\ \bibinfo {author} {\bibfnamefont {A.~M.}\
  \bibnamefont {Guloy}},\ }\bibfield  {title} {\bibinfo {title}
  {\ce{Ba$_{1–x}$Na$_x$Ti2Sb2O} ($0.0 \le x \le 0.33$): A layered
  titanium-based pnictide oxide superconductor},\ }\href@noop {} {\bibfield
  {journal} {\bibinfo  {journal} {J. Am. Chem. Soc.}\ }\textbf {\bibinfo
  {volume} {134}},\ \bibinfo {pages} {16520} (\bibinfo {year}
  {2012})}\BibitemShut {NoStop}%
\bibitem [{\citenamefont {Liu}\ \emph {et~al.}(2009)\citenamefont {Liu},
  \citenamefont {Tan}, \citenamefont {Song}, \citenamefont {Li}, \citenamefont
  {Yan}, \citenamefont {Ying}, \citenamefont {Xie}, \citenamefont {Wang},\ and\
  \citenamefont {Chen}}]{Liu2009}%
  \BibitemOpen
  \bibfield  {author} {\bibinfo {author} {\bibfnamefont {R.~H.}\ \bibnamefont
  {Liu}}, \bibinfo {author} {\bibfnamefont {D.}~\bibnamefont {Tan}}, \bibinfo
  {author} {\bibfnamefont {Y.~A.}\ \bibnamefont {Song}}, \bibinfo {author}
  {\bibfnamefont {Q.~J.}\ \bibnamefont {Li}}, \bibinfo {author} {\bibfnamefont
  {Y.~J.}\ \bibnamefont {Yan}}, \bibinfo {author} {\bibfnamefont {J.~J.}\
  \bibnamefont {Ying}}, \bibinfo {author} {\bibfnamefont {Y.~L.}\ \bibnamefont
  {Xie}}, \bibinfo {author} {\bibfnamefont {X.~F.}\ \bibnamefont {Wang}},\ and\
  \bibinfo {author} {\bibfnamefont {X.~H.}\ \bibnamefont {Chen}},\ }\bibfield
  {title} {\bibinfo {title} {Physical properties of the layered pnictide oxides
  \ce{Na2Ti2$P$2O} ({$P$}={As}, {Sb})},\ }\href@noop {} {\bibfield  {journal}
  {\bibinfo  {journal} {Phys. Rev. B}\ }\textbf {\bibinfo {volume} {80}},\
  \bibinfo {pages} {144516} (\bibinfo {year} {2009})}\BibitemShut {NoStop}%
\bibitem [{\citenamefont {Kitagawa}\ \emph {et~al.}(2013)\citenamefont
  {Kitagawa}, \citenamefont {Ishida}, \citenamefont {Nakano}, \citenamefont
  {Yajima},\ and\ \citenamefont {Kageyama}}]{Kitagawa2013}%
  \BibitemOpen
  \bibfield  {author} {\bibinfo {author} {\bibfnamefont {S.}~\bibnamefont
  {Kitagawa}}, \bibinfo {author} {\bibfnamefont {K.}~\bibnamefont {Ishida}},
  \bibinfo {author} {\bibfnamefont {K.}~\bibnamefont {Nakano}}, \bibinfo
  {author} {\bibfnamefont {T.}~\bibnamefont {Yajima}},\ and\ \bibinfo {author}
  {\bibfnamefont {H.}~\bibnamefont {Kageyama}},\ }\bibfield  {title} {\bibinfo
  {title} {$s$-wave superconductivity in superconducting \ce{BaTi2Sb2O}
  revealed by ${}^{121/123}${Sb}-{NMR}/nuclear quadrupole resonance
  measurements},\ }\href@noop {} {\bibfield  {journal} {\bibinfo  {journal}
  {Phys. Rev. B}\ }\textbf {\bibinfo {volume} {87}},\ \bibinfo {pages}
  {060510(R)} (\bibinfo {year} {2013})}\BibitemShut {NoStop}%
\bibitem [{\citenamefont {von Rohr}\ \emph {et~al.}(2013)\citenamefont {von
  Rohr}, \citenamefont {Schilling}, \citenamefont {Nesper}, \citenamefont
  {Baines},\ and\ \citenamefont {Bendele}}]{vonRohr2013}%
  \BibitemOpen
  \bibfield  {author} {\bibinfo {author} {\bibfnamefont {F.}~\bibnamefont {von
  Rohr}}, \bibinfo {author} {\bibfnamefont {A.}~\bibnamefont {Schilling}},
  \bibinfo {author} {\bibfnamefont {R.}~\bibnamefont {Nesper}}, \bibinfo
  {author} {\bibfnamefont {C.}~\bibnamefont {Baines}},\ and\ \bibinfo {author}
  {\bibfnamefont {M.}~\bibnamefont {Bendele}},\ }\bibfield  {title} {\bibinfo
  {title} {Conventional superconductivity and charge-density-wave ordering in
  \ce{Ba$_{1-x}$Na$_x$Ti2Sb2O}},\ }\href@noop {} {\bibfield  {journal}
  {\bibinfo  {journal} {Phys. Rev. B}\ }\textbf {\bibinfo {volume} {88}},\
  \bibinfo {pages} {140501(R)} (\bibinfo {year} {2013})}\BibitemShut {NoStop}%
\bibitem [{\citenamefont {Nozaki}\ \emph {et~al.}(2013)\citenamefont {Nozaki},
  \citenamefont {Nakano}, \citenamefont {Yajima}, \citenamefont {Kageyama},
  \citenamefont {Frandsen}, \citenamefont {Liu}, \citenamefont {Cheung},
  \citenamefont {Goko}, \citenamefont {Uemura}, \citenamefont {Munsie},
  \citenamefont {Medina}, \citenamefont {Luke}, \citenamefont {Munevar},
  \citenamefont {Nishio-Hamane},\ and\ \citenamefont {Brown}}]{Nozaki2013}%
  \BibitemOpen
  \bibfield  {author} {\bibinfo {author} {\bibfnamefont {Y.}~\bibnamefont
  {Nozaki}}, \bibinfo {author} {\bibfnamefont {K.}~\bibnamefont {Nakano}},
  \bibinfo {author} {\bibfnamefont {T.}~\bibnamefont {Yajima}}, \bibinfo
  {author} {\bibfnamefont {H.}~\bibnamefont {Kageyama}}, \bibinfo {author}
  {\bibfnamefont {B.}~\bibnamefont {Frandsen}}, \bibinfo {author}
  {\bibfnamefont {L.}~\bibnamefont {Liu}}, \bibinfo {author} {\bibfnamefont
  {S.}~\bibnamefont {Cheung}}, \bibinfo {author} {\bibfnamefont
  {T.}~\bibnamefont {Goko}}, \bibinfo {author} {\bibfnamefont {Y.~J.}\
  \bibnamefont {Uemura}}, \bibinfo {author} {\bibfnamefont {T.~S.~J.}\
  \bibnamefont {Munsie}}, \bibinfo {author} {\bibfnamefont {T.}~\bibnamefont
  {Medina}}, \bibinfo {author} {\bibfnamefont {G.~M.}\ \bibnamefont {Luke}},
  \bibinfo {author} {\bibfnamefont {J.}~\bibnamefont {Munevar}}, \bibinfo
  {author} {\bibfnamefont {D.}~\bibnamefont {Nishio-Hamane}},\ and\ \bibinfo
  {author} {\bibfnamefont {C.~M.}\ \bibnamefont {Brown}},\ }\bibfield  {title}
  {\bibinfo {title} {Muon spin relaxation and electron/neutron diffraction
  studies of \ce{BaTi2(As$_{1-x}$Sb$_x$)2O}: Absence of static magnetism and
  superlattice reflections},\ }\href@noop {} {\bibfield  {journal} {\bibinfo
  {journal} {Phys. Rev. B}\ }\textbf {\bibinfo {volume} {88}},\ \bibinfo
  {pages} {214506} (\bibinfo {year} {2013})}\BibitemShut {NoStop}%
\bibitem [{\citenamefont {Pachmayr}\ and\ \citenamefont
  {Johrendt}(2014)}]{Pachmayr2014}%
  \BibitemOpen
  \bibfield  {author} {\bibinfo {author} {\bibfnamefont {U.}~\bibnamefont
  {Pachmayr}}\ and\ \bibinfo {author} {\bibfnamefont {D.}~\bibnamefont
  {Johrendt}},\ }\bibfield  {title} {\bibinfo {title} {Superconductivity in
  \ce{Ba$_{1-x}$K$_x$Ti2Sb2O} ($0\le x \le 1$) controlled by the layer
  charge},\ }\href@noop {} {\bibfield  {journal} {\bibinfo  {journal} {Solid
  State Sci.}\ }\textbf {\bibinfo {volume} {28}},\ \bibinfo {pages} {31}
  (\bibinfo {year} {2014})}\BibitemShut {NoStop}%
\bibitem [{\citenamefont {von Rohr}\ \emph {et~al.}(2014)\citenamefont {von
  Rohr}, \citenamefont {Nesper},\ and\ \citenamefont
  {Schilling}}]{vonRohr2014}%
  \BibitemOpen
  \bibfield  {author} {\bibinfo {author} {\bibfnamefont {F.}~\bibnamefont {von
  Rohr}}, \bibinfo {author} {\bibfnamefont {R.}~\bibnamefont {Nesper}},\ and\
  \bibinfo {author} {\bibfnamefont {A.}~\bibnamefont {Schilling}},\ }\bibfield
  {title} {\bibinfo {title} {Superconductivity in rubidium-substituted
  \ce{Ba$_{1-x}$Rb$_x$Ti2Sb2O}},\ }\href@noop {} {\bibfield  {journal}
  {\bibinfo  {journal} {Phys. Rev. B}\ }\textbf {\bibinfo {volume} {89}},\
  \bibinfo {pages} {094505} (\bibinfo {year} {2014})}\BibitemShut {NoStop}%
\bibitem [{\citenamefont {Wang}\ \emph {et~al.}(2019)\citenamefont {Wang},
  \citenamefont {Yang}, \citenamefont {Taguchi}, \citenamefont {Li},
  \citenamefont {He}, \citenamefont {Goto}, \citenamefont {Eguchi},
  \citenamefont {Miyazaki}, \citenamefont {Liao}, \citenamefont {Ishii},\ and\
  \citenamefont {Kubozono}}]{Wang2019}%
  \BibitemOpen
  \bibfield  {author} {\bibinfo {author} {\bibfnamefont {Y.}~\bibnamefont
  {Wang}}, \bibinfo {author} {\bibfnamefont {X.}~\bibnamefont {Yang}}, \bibinfo
  {author} {\bibfnamefont {T.}~\bibnamefont {Taguchi}}, \bibinfo {author}
  {\bibfnamefont {H.}~\bibnamefont {Li}}, \bibinfo {author} {\bibfnamefont
  {T.}~\bibnamefont {He}}, \bibinfo {author} {\bibfnamefont {H.}~\bibnamefont
  {Goto}}, \bibinfo {author} {\bibfnamefont {R.}~\bibnamefont {Eguchi}},
  \bibinfo {author} {\bibfnamefont {T.}~\bibnamefont {Miyazaki}}, \bibinfo
  {author} {\bibfnamefont {Y.-F.}\ \bibnamefont {Liao}}, \bibinfo {author}
  {\bibfnamefont {H.}~\bibnamefont {Ishii}},\ and\ \bibinfo {author}
  {\bibfnamefont {Y.}~\bibnamefont {Kubozono}},\ }\bibfield  {title} {\bibinfo
  {title} {Preparation and characterization of superconducting
  \ce{Ba$_{1-x}$Cs$_x$Ti2Sb2O}, and its pressure dependence of
  superconductivity},\ }\href@noop {} {\bibfield  {journal} {\bibinfo
  {journal} {Jpn. J. Appl. Phys.}\ }\textbf {\bibinfo {volume} {58}},\ \bibinfo
  {pages} {110603} (\bibinfo {year} {2019})}\BibitemShut {NoStop}%
\bibitem [{\citenamefont {Ishii}\ \emph {et~al.}(2018)\citenamefont {Ishii},
  \citenamefont {Yajima},\ and\ \citenamefont {Hiroi}}]{Ishii2018}%
  \BibitemOpen
  \bibfield  {author} {\bibinfo {author} {\bibfnamefont {W.}~\bibnamefont
  {Ishii}}, \bibinfo {author} {\bibfnamefont {T.}~\bibnamefont {Yajima}},\ and\
  \bibinfo {author} {\bibfnamefont {Z.}~\bibnamefont {Hiroi}},\ }\bibfield
  {title} {\bibinfo {title} {Electronic phase diagram of the titanium
  oxypnictide superconductor \ce{BaTi2(Sb$_{1-x}$Bi$_x$)2O}},\ }\href@noop {}
  {\bibfield  {journal} {\bibinfo  {journal} {J. Phys.: Conf. Ser.}\ }\textbf
  {\bibinfo {volume} {969}},\ \bibinfo {pages} {012052} (\bibinfo {year}
  {2018})}\BibitemShut {NoStop}%
\bibitem [{\citenamefont {Wang}\ \emph {et~al.}(2020)\citenamefont {Wang},
  \citenamefont {Li}, \citenamefont {Taguchi}, \citenamefont {Suzuki},
  \citenamefont {Miura}, \citenamefont {Goto}, \citenamefont {Eguchi},
  \citenamefont {Miyazaki}, \citenamefont {Liao}, \citenamefont {Ishii},\ and\
  \citenamefont {Kubozono}}]{Wang2020}%
  \BibitemOpen
  \bibfield  {author} {\bibinfo {author} {\bibfnamefont {Y.}~\bibnamefont
  {Wang}}, \bibinfo {author} {\bibfnamefont {H.}~\bibnamefont {Li}}, \bibinfo
  {author} {\bibfnamefont {T.}~\bibnamefont {Taguchi}}, \bibinfo {author}
  {\bibfnamefont {A.}~\bibnamefont {Suzuki}}, \bibinfo {author} {\bibfnamefont
  {A.}~\bibnamefont {Miura}}, \bibinfo {author} {\bibfnamefont
  {H.}~\bibnamefont {Goto}}, \bibinfo {author} {\bibfnamefont {R.}~\bibnamefont
  {Eguchi}}, \bibinfo {author} {\bibfnamefont {T.}~\bibnamefont {Miyazaki}},
  \bibinfo {author} {\bibfnamefont {Y.-F.}\ \bibnamefont {Liao}}, \bibinfo
  {author} {\bibfnamefont {H.}~\bibnamefont {Ishii}},\ and\ \bibinfo {author}
  {\bibfnamefont {Y.}~\bibnamefont {Kubozono}},\ }\bibfield  {title} {\bibinfo
  {title} {Superconducting behavior of \ce{BaTi2Bi2O} and its pressure
  dependence},\ }\href@noop {} {\bibfield  {journal} {\bibinfo  {journal}
  {Phys. Chem. Chem. Phys.}\ }\textbf {\bibinfo {volume} {22}},\ \bibinfo
  {pages} {23315} (\bibinfo {year} {2020})}\BibitemShut {NoStop}%
\bibitem [{\citenamefont {Lorenz}\ \emph {et~al.}(2014)\citenamefont {Lorenz},
  \citenamefont {Guloy},\ and\ \citenamefont {Chu}}]{Lorenz2014}%
  \BibitemOpen
  \bibfield  {author} {\bibinfo {author} {\bibfnamefont {B.}~\bibnamefont
  {Lorenz}}, \bibinfo {author} {\bibfnamefont {A.~M.}\ \bibnamefont {Guloy}},\
  and\ \bibinfo {author} {\bibfnamefont {P.~C.~W.}\ \bibnamefont {Chu}},\
  }\bibfield  {title} {\bibinfo {title} {Superconductivity in titanium-based
  pnictide oxide compounds},\ }\href
  {https://doi.org/10.1142/S0217979214300114} {\bibfield  {journal} {\bibinfo
  {journal} {Int. J. Mod. Phys. B}\ }\textbf {\bibinfo {volume} {28}},\
  \bibinfo {pages} {1430011} (\bibinfo {year} {2014})}\BibitemShut {NoStop}%
\bibitem [{\citenamefont {Gooch}\ \emph {et~al.}(2013)\citenamefont {Gooch},
  \citenamefont {Doan}, \citenamefont {Tang}, \citenamefont {Lorenz},
  \citenamefont {Guloy},\ and\ \citenamefont {Chu}}]{Gooch2013}%
  \BibitemOpen
  \bibfield  {author} {\bibinfo {author} {\bibfnamefont {M.}~\bibnamefont
  {Gooch}}, \bibinfo {author} {\bibfnamefont {P.}~\bibnamefont {Doan}},
  \bibinfo {author} {\bibfnamefont {Z.}~\bibnamefont {Tang}}, \bibinfo {author}
  {\bibfnamefont {B.}~\bibnamefont {Lorenz}}, \bibinfo {author} {\bibfnamefont
  {A.~M.}\ \bibnamefont {Guloy}},\ and\ \bibinfo {author} {\bibfnamefont
  {P.~C.~W.}\ \bibnamefont {Chu}},\ }\bibfield  {title} {\bibinfo {title} {Weak
  coupling {BCS}-like superconductivity in the pnictide oxide
  \ce{Ba$_{1-x}$Na$_x$Ti2Sb2O} ($x=0$ and 0.15)},\ }\href@noop {} {\bibfield
  {journal} {\bibinfo  {journal} {Phys. Rev. B}\ }\textbf {\bibinfo {volume}
  {88}},\ \bibinfo {pages} {064510} (\bibinfo {year} {2013})}\BibitemShut
  {NoStop}%
\bibitem [{\citenamefont {Kamusella}\ \emph {et~al.}(2014)\citenamefont
  {Kamusella}, \citenamefont {Doan}, \citenamefont {Goltz}, \citenamefont
  {Luetkens}, \citenamefont {Sarkar}, \citenamefont {Guloy},\ and\
  \citenamefont {Klauss}}]{Kamusella2014}%
  \BibitemOpen
  \bibfield  {author} {\bibinfo {author} {\bibfnamefont {S.}~\bibnamefont
  {Kamusella}}, \bibinfo {author} {\bibfnamefont {P.}~\bibnamefont {Doan}},
  \bibinfo {author} {\bibfnamefont {T.}~\bibnamefont {Goltz}}, \bibinfo
  {author} {\bibfnamefont {H.}~\bibnamefont {Luetkens}}, \bibinfo {author}
  {\bibfnamefont {R.}~\bibnamefont {Sarkar}}, \bibinfo {author} {\bibfnamefont
  {A.}~\bibnamefont {Guloy}},\ and\ \bibinfo {author} {\bibfnamefont {H.-H.}\
  \bibnamefont {Klauss}},\ }\bibfield  {title} {\bibinfo {title} {{CDW} order
  and unconventional s-wave superconductivity in
  \ce{Ba$_{1-x}$Na$_x$Ti2Sb2O}},\ }\href@noop {} {\bibfield  {journal}
  {\bibinfo  {journal} {J. Phys.: Conf. Ser.}\ }\textbf {\bibinfo {volume}
  {551}},\ \bibinfo {pages} {012026} (\bibinfo {year} {2014})}\BibitemShut
  {NoStop}%
\bibitem [{\citenamefont {Kitagawa}\ \emph {et~al.}(2018)\citenamefont
  {Kitagawa}, \citenamefont {Ishida}, \citenamefont {Ishii}, \citenamefont
  {Yajima},\ and\ \citenamefont {Hiroi}}]{Kitagawa2018}%
  \BibitemOpen
  \bibfield  {author} {\bibinfo {author} {\bibfnamefont {S.}~\bibnamefont
  {Kitagawa}}, \bibinfo {author} {\bibfnamefont {K.}~\bibnamefont {Ishida}},
  \bibinfo {author} {\bibfnamefont {W.}~\bibnamefont {Ishii}}, \bibinfo
  {author} {\bibfnamefont {T.}~\bibnamefont {Yajima}},\ and\ \bibinfo {author}
  {\bibfnamefont {Z.}~\bibnamefont {Hiroi}},\ }\bibfield  {title} {\bibinfo
  {title} {Nematic transition and highly two-dimensional superconductivity in
  \ce{BaTi2Bi2O} revealed by $^{209}${Bi}-nuclear magnetic resonance/nuclear
  quadrupole resonance measurements},\ }\href@noop {} {\bibfield  {journal}
  {\bibinfo  {journal} {Phys. Rev. B}\ }\textbf {\bibinfo {volume} {98}},\
  \bibinfo {pages} {220507(R)} (\bibinfo {year} {2018})}\BibitemShut {NoStop}%
\bibitem [{\citenamefont {Subedi}(2013)}]{Subedi2013}%
  \BibitemOpen
  \bibfield  {author} {\bibinfo {author} {\bibfnamefont {A.}~\bibnamefont
  {Subedi}},\ }\bibfield  {title} {\bibinfo {title} {Electron-phonon
  superconductivity and charge density wave instability in the layered
  titanium-based pnictide \ce{BaTi2Sb2O}},\ }\href@noop {} {\bibfield
  {journal} {\bibinfo  {journal} {Phys. Rev. B}\ }\textbf {\bibinfo {volume}
  {87}},\ \bibinfo {pages} {054506} (\bibinfo {year} {2013})}\BibitemShut
  {NoStop}%
\bibitem [{\citenamefont {Singh}(2012)}]{Singh2012}%
  \BibitemOpen
  \bibfield  {author} {\bibinfo {author} {\bibfnamefont {D.~J.}\ \bibnamefont
  {Singh}},\ }\bibfield  {title} {\bibinfo {title} {Electronic structure,
  disconnected {Fermi} surfaces and antiferromagnetism in the layered pnictide
  superconductor \ce{Na$_x$Ba$_{1-x}$Ti2Sb2O}},\ }\href
  {https://doi.org/10.1088/1367-2630/14/12/123003} {\bibfield  {journal}
  {\bibinfo  {journal} {New J. Phys.}\ }\textbf {\bibinfo {volume} {14}},\
  \bibinfo {pages} {123003} (\bibinfo {year} {2012})}\BibitemShut {NoStop}%
\bibitem [{\citenamefont {Zhang}\ \emph {et~al.}(2017)\citenamefont {Zhang},
  \citenamefont {Glasbrenner}, \citenamefont {Flint}, \citenamefont {Mazin},\
  and\ \citenamefont {Fernandes}}]{Zhang2017}%
  \BibitemOpen
  \bibfield  {author} {\bibinfo {author} {\bibfnamefont {G.}~\bibnamefont
  {Zhang}}, \bibinfo {author} {\bibfnamefont {J.~K.}\ \bibnamefont
  {Glasbrenner}}, \bibinfo {author} {\bibfnamefont {R.}~\bibnamefont {Flint}},
  \bibinfo {author} {\bibfnamefont {I.~I.}\ \bibnamefont {Mazin}},\ and\
  \bibinfo {author} {\bibfnamefont {R.~M.}\ \bibnamefont {Fernandes}},\
  }\bibfield  {title} {\bibinfo {title} {Double-stage nematic bond ordering
  above double stripe magnetism: Application to \ce{BaTi2Sb2O}},\ }\href@noop
  {} {\bibfield  {journal} {\bibinfo  {journal} {Phys. Rev. B}\ }\textbf
  {\bibinfo {volume} {95}},\ \bibinfo {pages} {174402} (\bibinfo {year}
  {2017})}\BibitemShut {NoStop}%
\bibitem [{\citenamefont {Yu}\ \emph {et~al.}(2014)\citenamefont {Yu},
  \citenamefont {Liu}, \citenamefont {Quan}, \citenamefont {Jia}, \citenamefont
  {Lin},\ and\ \citenamefont {Zou}}]{Yu2014}%
  \BibitemOpen
  \bibfield  {author} {\bibinfo {author} {\bibfnamefont {X.-L.}\ \bibnamefont
  {Yu}}, \bibinfo {author} {\bibfnamefont {D.-Y.}\ \bibnamefont {Liu}},
  \bibinfo {author} {\bibfnamefont {Y.-M.}\ \bibnamefont {Quan}}, \bibinfo
  {author} {\bibfnamefont {T.}~\bibnamefont {Jia}}, \bibinfo {author}
  {\bibfnamefont {H.-Q.}\ \bibnamefont {Lin}},\ and\ \bibinfo {author}
  {\bibfnamefont {L.-J.}\ \bibnamefont {Zou}},\ }\bibfield  {title} {\bibinfo
  {title} {A site-selective antiferromagnetic ground state in layered
  pnictide-oxide \ce{BaTi2As2O}},\ }\href@noop {} {\bibfield  {journal}
  {\bibinfo  {journal} {J. Appl. Phys.}\ }\textbf {\bibinfo {volume} {115}},\
  \bibinfo {pages} {17A924} (\bibinfo {year} {2014})}\BibitemShut {NoStop}%
\bibitem [{\citenamefont {Kim}\ \emph {et~al.}(2017)\citenamefont {Kim},
  \citenamefont {Shim}, \citenamefont {Kim},\ and\ \citenamefont
  {Min}}]{Kim2017}%
  \BibitemOpen
  \bibfield  {author} {\bibinfo {author} {\bibfnamefont {H.}~\bibnamefont
  {Kim}}, \bibinfo {author} {\bibfnamefont {J.~H.}\ \bibnamefont {Shim}},
  \bibinfo {author} {\bibfnamefont {K.}~\bibnamefont {Kim}},\ and\ \bibinfo
  {author} {\bibfnamefont {B.~I.}\ \bibnamefont {Min}},\ }\bibfield  {title}
  {\bibinfo {title} {Charge density waves and the {Coulomb} correlation effects
  in \ce{Na2Ti2$P$2O} ({$P$}={Sb}, {As})},\ }\href@noop {} {\bibfield
  {journal} {\bibinfo  {journal} {Phys. Rev. B}\ }\textbf {\bibinfo {volume}
  {96}},\ \bibinfo {pages} {155142} (\bibinfo {year} {2017})}\BibitemShut
  {NoStop}%
\bibitem [{\citenamefont {Wang}\ \emph {et~al.}(2013)\citenamefont {Wang},
  \citenamefont {Zhang}, \citenamefont {Zhang},\ and\ \citenamefont
  {Liu}}]{Wang2013}%
  \BibitemOpen
  \bibfield  {author} {\bibinfo {author} {\bibfnamefont {G.}~\bibnamefont
  {Wang}}, \bibinfo {author} {\bibfnamefont {H.}~\bibnamefont {Zhang}},
  \bibinfo {author} {\bibfnamefont {L.}~\bibnamefont {Zhang}},\ and\ \bibinfo
  {author} {\bibfnamefont {C.}~\bibnamefont {Liu}},\ }\bibfield  {title}
  {\bibinfo {title} {The electronic structure and magnetism of
  \ce{BaTi2Sb2O}},\ }\href@noop {} {\bibfield  {journal} {\bibinfo  {journal}
  {J. Appl. Phys.}\ }\textbf {\bibinfo {volume} {113}},\ \bibinfo {pages}
  {243904} (\bibinfo {year} {2013})}\BibitemShut {NoStop}%
\bibitem [{\citenamefont {Yan}\ and\ \citenamefont {Lu}(2013)}]{Yan2013}%
  \BibitemOpen
  \bibfield  {author} {\bibinfo {author} {\bibfnamefont {X.-W.}\ \bibnamefont
  {Yan}}\ and\ \bibinfo {author} {\bibfnamefont {Z.-Y.}\ \bibnamefont {Lu}},\
  }\bibfield  {title} {\bibinfo {title} {Layered pnictide-oxide
  \ce{Na2Ti2$Pn$2O} ({$Pn$}={As}, {Sb}): a candidate for spin density waves},\
  }\href@noop {} {\bibfield  {journal} {\bibinfo  {journal} {J. Phys.: Condens.
  Matter}\ }\textbf {\bibinfo {volume} {25}},\ \bibinfo {pages} {365501}
  (\bibinfo {year} {2013})}\BibitemShut {NoStop}%
\bibitem [{\citenamefont {Koepernik}\ and\ \citenamefont
  {Eschrig}(1999)}]{Koepernik1999}%
  \BibitemOpen
  \bibfield  {author} {\bibinfo {author} {\bibfnamefont {K.}~\bibnamefont
  {Koepernik}}\ and\ \bibinfo {author} {\bibfnamefont {H.}~\bibnamefont
  {Eschrig}},\ }\bibfield  {title} {\bibinfo {title} {Full-potential
  nonorthogonal local-orbital minimum-basis band-structure scheme},\
  }\href@noop {} {\bibfield  {journal} {\bibinfo  {journal} {Phys. Rev. B}\
  }\textbf {\bibinfo {volume} {59}},\ \bibinfo {pages} {1743} (\bibinfo {year}
  {1999})}\BibitemShut {NoStop}%
\bibitem [{\citenamefont {Perdew}\ \emph {et~al.}(1996)\citenamefont {Perdew},
  \citenamefont {Burke},\ and\ \citenamefont {Ernzerhof}}]{Perdew1996}%
  \BibitemOpen
  \bibfield  {author} {\bibinfo {author} {\bibfnamefont {J.~P.}\ \bibnamefont
  {Perdew}}, \bibinfo {author} {\bibfnamefont {K.}~\bibnamefont {Burke}},\ and\
  \bibinfo {author} {\bibfnamefont {M.}~\bibnamefont {Ernzerhof}},\ }\bibfield
  {title} {\bibinfo {title} {Generalized gradient approximation made simple},\
  }\href {https://doi.org/10.1103/PhysRevLett.77.3865} {\bibfield  {journal}
  {\bibinfo  {journal} {Phys. Rev. Lett.}\ }\textbf {\bibinfo {volume} {77}},\
  \bibinfo {pages} {3865} (\bibinfo {year} {1996})}\BibitemShut {NoStop}%
\bibitem [{\citenamefont {Eschrig}\ and\ \citenamefont
  {Koepernik}(2009)}]{Eschrig2009}%
  \BibitemOpen
  \bibfield  {author} {\bibinfo {author} {\bibfnamefont {H.}~\bibnamefont
  {Eschrig}}\ and\ \bibinfo {author} {\bibfnamefont {K.}~\bibnamefont
  {Koepernik}},\ }\bibfield  {title} {\bibinfo {title} {Tight-binding models
  for the iron-based superconductors},\ }\href@noop {} {\bibfield  {journal}
  {\bibinfo  {journal} {Phys. Rev. B}\ }\textbf {\bibinfo {volume} {80}},\
  \bibinfo {pages} {104503} (\bibinfo {year} {2009})}\BibitemShut {NoStop}%
\bibitem [{\citenamefont {Guterding}\ \emph {et~al.}(2015)\citenamefont
  {Guterding}, \citenamefont {Jeschke}, \citenamefont {Hirschfeld},\ and\
  \citenamefont {Valent\'{\i}}}]{Guterding2015}%
  \BibitemOpen
  \bibfield  {author} {\bibinfo {author} {\bibfnamefont {D.}~\bibnamefont
  {Guterding}}, \bibinfo {author} {\bibfnamefont {H.~O.}\ \bibnamefont
  {Jeschke}}, \bibinfo {author} {\bibfnamefont {P.~J.}\ \bibnamefont
  {Hirschfeld}},\ and\ \bibinfo {author} {\bibfnamefont {R.}~\bibnamefont
  {Valent\'{\i}}},\ }\bibfield  {title} {\bibinfo {title} {Unified picture of
  the doping dependence of superconducting transition temperatures in alkali
  metal/ammonia intercalated {FeSe}},\ }\href@noop {} {\bibfield  {journal}
  {\bibinfo  {journal} {Phys. Rev. B}\ }\textbf {\bibinfo {volume} {91}},\
  \bibinfo {pages} {041112(R)} (\bibinfo {year} {2015})}\BibitemShut {NoStop}%
\bibitem [{\citenamefont {Guterding}(2017)}]{Guterding2017}%
  \BibitemOpen
  \bibfield  {author} {\bibinfo {author} {\bibfnamefont {D.}~\bibnamefont
  {Guterding}},\ }\emph {\bibinfo {title} {Microscopic modelling of organic and
  iron-based superconductors}},\ \href@noop {} {Ph.D. thesis},\ \bibinfo
  {school} {Goethe-Universit{\"a}t Frankfurt, Germany} (\bibinfo {year}
  {2017})\BibitemShut {NoStop}%
\bibitem [{\citenamefont {Shimizu}\ \emph {et~al.}(2018)\citenamefont
  {Shimizu}, \citenamefont {Takemori}, \citenamefont {Guterding},\ and\
  \citenamefont {Jeschke}}]{Shimizu2018}%
  \BibitemOpen
  \bibfield  {author} {\bibinfo {author} {\bibfnamefont {M.}~\bibnamefont
  {Shimizu}}, \bibinfo {author} {\bibfnamefont {N.}~\bibnamefont {Takemori}},
  \bibinfo {author} {\bibfnamefont {D.}~\bibnamefont {Guterding}},\ and\
  \bibinfo {author} {\bibfnamefont {H.~O.}\ \bibnamefont {Jeschke}},\
  }\bibfield  {title} {\bibinfo {title} {Two-dome superconductivity in {FeS}
  induced by a {Lifshitz} transition},\ }\href@noop {} {\bibfield  {journal}
  {\bibinfo  {journal} {Phys. Rev. Lett.}\ }\textbf {\bibinfo {volume} {121}},\
  \bibinfo {pages} {137001} (\bibinfo {year} {2018})}\BibitemShut {NoStop}%
\bibitem [{\citenamefont {Shimizu}\ \emph {et~al.}(2020)\citenamefont
  {Shimizu}, \citenamefont {Takemori}, \citenamefont {Guterding},\ and\
  \citenamefont {Jeschke}}]{Shimizu2020}%
  \BibitemOpen
  \bibfield  {author} {\bibinfo {author} {\bibfnamefont {M.}~\bibnamefont
  {Shimizu}}, \bibinfo {author} {\bibfnamefont {N.}~\bibnamefont {Takemori}},
  \bibinfo {author} {\bibfnamefont {D.}~\bibnamefont {Guterding}},\ and\
  \bibinfo {author} {\bibfnamefont {H.~O.}\ \bibnamefont {Jeschke}},\
  }\bibfield  {title} {\bibinfo {title} {Importance of the {Fermi} surface and
  magnetic interactions for the superconducting dome in electron-doped {FeSe}
  intercalates},\ }\href@noop {} {\bibfield  {journal} {\bibinfo  {journal}
  {Phys. Rev. B}\ }\textbf {\bibinfo {volume} {101}},\ \bibinfo {pages}
  {180511(R)} (\bibinfo {year} {2020})}\BibitemShut {NoStop}%
\bibitem [{\citenamefont {Hosono}\ \emph {et~al.}(2015)\citenamefont {Hosono},
  \citenamefont {Tanabe}, \citenamefont {Takayama-Muromachi}, \citenamefont
  {Kageyama}, \citenamefont {Yamanaka}, \citenamefont {Kumakura}, \citenamefont
  {Nohara}, \citenamefont {Hiramatsu},\ and\ \citenamefont
  {Fujitsu}}]{Hosono2015}%
  \BibitemOpen
  \bibfield  {author} {\bibinfo {author} {\bibfnamefont {H.}~\bibnamefont
  {Hosono}}, \bibinfo {author} {\bibfnamefont {K.}~\bibnamefont {Tanabe}},
  \bibinfo {author} {\bibfnamefont {E.}~\bibnamefont {Takayama-Muromachi}},
  \bibinfo {author} {\bibfnamefont {H.}~\bibnamefont {Kageyama}}, \bibinfo
  {author} {\bibfnamefont {S.}~\bibnamefont {Yamanaka}}, \bibinfo {author}
  {\bibfnamefont {H.}~\bibnamefont {Kumakura}}, \bibinfo {author}
  {\bibfnamefont {M.}~\bibnamefont {Nohara}}, \bibinfo {author} {\bibfnamefont
  {H.}~\bibnamefont {Hiramatsu}},\ and\ \bibinfo {author} {\bibfnamefont
  {S.}~\bibnamefont {Fujitsu}},\ }\bibfield  {title} {\bibinfo {title}
  {Exploration of new superconductors and functional materials, and fabrication
  of superconducting tapes and wires of iron pnictides},\ }\href
  {https://doi.org/10.1088/1468-6996/16/3/033503} {\bibfield  {journal}
  {\bibinfo  {journal} {Sci. Technol. Adv. Mater.}\ }\textbf {\bibinfo {volume}
  {16}},\ \bibinfo {pages} {033503} (\bibinfo {year} {2015})}\BibitemShut
  {NoStop}%
\bibitem [{\citenamefont {Nakaoka}\ \emph {et~al.}(2016)\citenamefont
  {Nakaoka}, \citenamefont {Yamakawa},\ and\ \citenamefont
  {Kontani}}]{Nakaoka2016}%
  \BibitemOpen
  \bibfield  {author} {\bibinfo {author} {\bibfnamefont {H.}~\bibnamefont
  {Nakaoka}}, \bibinfo {author} {\bibfnamefont {Y.}~\bibnamefont {Yamakawa}},\
  and\ \bibinfo {author} {\bibfnamefont {H.}~\bibnamefont {Kontani}},\
  }\bibfield  {title} {\bibinfo {title} {Theoretical prediction of nematic
  orbital-ordered state in the {Ti} oxypnictide superconductor
  \ce{BaTi2(As,Sb)2O}},\ }\href@noop {} {\bibfield  {journal} {\bibinfo
  {journal} {Phys. Rev. B}\ }\textbf {\bibinfo {volume} {93}},\ \bibinfo
  {pages} {245122} (\bibinfo {year} {2016})}\BibitemShut {NoStop}%
\bibitem [{\citenamefont {Song}\ \emph {et~al.}(2018)\citenamefont {Song},
  \citenamefont {Li}, \citenamefont {Zhao}, \citenamefont {Ma}, \citenamefont
  {Zheng}, \citenamefont {Li}, \citenamefont {Nie}, \citenamefont {Luo},
  \citenamefont {Yin}, \citenamefont {Wu},\ and\ \citenamefont
  {Chen}}]{Song2018}%
  \BibitemOpen
  \bibfield  {author} {\bibinfo {author} {\bibfnamefont {D.~W.}\ \bibnamefont
  {Song}}, \bibinfo {author} {\bibfnamefont {J.}~\bibnamefont {Li}}, \bibinfo
  {author} {\bibfnamefont {D.}~\bibnamefont {Zhao}}, \bibinfo {author}
  {\bibfnamefont {L.~K.}\ \bibnamefont {Ma}}, \bibinfo {author} {\bibfnamefont
  {L.~X.}\ \bibnamefont {Zheng}}, \bibinfo {author} {\bibfnamefont {S.~J.}\
  \bibnamefont {Li}}, \bibinfo {author} {\bibfnamefont {L.~P.}\ \bibnamefont
  {Nie}}, \bibinfo {author} {\bibfnamefont {X.~G.}\ \bibnamefont {Luo}},
  \bibinfo {author} {\bibfnamefont {Z.~P.}\ \bibnamefont {Yin}}, \bibinfo
  {author} {\bibfnamefont {T.}~\bibnamefont {Wu}},\ and\ \bibinfo {author}
  {\bibfnamefont {X.~H.}\ \bibnamefont {Chen}},\ }\bibfield  {title} {\bibinfo
  {title} {Revealing the hidden order in \ce{BaTi2As2O} via nuclear magnetic
  resonance},\ }\href@noop {} {\bibfield  {journal} {\bibinfo  {journal} {Phys.
  Rev. B}\ }\textbf {\bibinfo {volume} {98}},\ \bibinfo {pages} {235142}
  (\bibinfo {year} {2018})}\BibitemShut {NoStop}%
\bibitem [{\citenamefont {Davies}\ \emph {et~al.}(2016)\citenamefont {Davies},
  \citenamefont {Johnson}, \citenamefont {Princep}, \citenamefont {Gannon},
  \citenamefont {Ma}, \citenamefont {Qian}, \citenamefont {Richard},
  \citenamefont {Li}, \citenamefont {Shi}, \citenamefont {Nowell},
  \citenamefont {Baker}, \citenamefont {Shi}, \citenamefont {Ding},
  \citenamefont {Luo}, \citenamefont {Guo},\ and\ \citenamefont
  {Boothroyd}}]{Davies2016}%
  \BibitemOpen
  \bibfield  {author} {\bibinfo {author} {\bibfnamefont {N.~R.}\ \bibnamefont
  {Davies}}, \bibinfo {author} {\bibfnamefont {R.~D.}\ \bibnamefont {Johnson}},
  \bibinfo {author} {\bibfnamefont {A.~J.}\ \bibnamefont {Princep}}, \bibinfo
  {author} {\bibfnamefont {L.~A.}\ \bibnamefont {Gannon}}, \bibinfo {author}
  {\bibfnamefont {J.-Z.}\ \bibnamefont {Ma}}, \bibinfo {author} {\bibfnamefont
  {T.}~\bibnamefont {Qian}}, \bibinfo {author} {\bibfnamefont {P.}~\bibnamefont
  {Richard}}, \bibinfo {author} {\bibfnamefont {H.}~\bibnamefont {Li}},
  \bibinfo {author} {\bibfnamefont {M.}~\bibnamefont {Shi}}, \bibinfo {author}
  {\bibfnamefont {H.}~\bibnamefont {Nowell}}, \bibinfo {author} {\bibfnamefont
  {P.~J.}\ \bibnamefont {Baker}}, \bibinfo {author} {\bibfnamefont {Y.~G.}\
  \bibnamefont {Shi}}, \bibinfo {author} {\bibfnamefont {H.}~\bibnamefont
  {Ding}}, \bibinfo {author} {\bibfnamefont {J.}~\bibnamefont {Luo}}, \bibinfo
  {author} {\bibfnamefont {Y.~F.}\ \bibnamefont {Guo}},\ and\ \bibinfo {author}
  {\bibfnamefont {A.~T.}\ \bibnamefont {Boothroyd}},\ }\bibfield  {title}
  {\bibinfo {title} {Coupled commensurate charge density wave and lattice
  distortion in \ce{Na2Ti2$Pn$2O} ({$Pn$}={As},{Sb}) determined by x-ray
  diffraction and angle-resolved photoemission spectroscopy},\ }\href@noop {}
  {\bibfield  {journal} {\bibinfo  {journal} {Phys. Rev. B}\ }\textbf {\bibinfo
  {volume} {94}},\ \bibinfo {pages} {104515} (\bibinfo {year}
  {2016})}\BibitemShut {NoStop}%
\bibitem [{\citenamefont {Huang}\ \emph {et~al.}(2020)\citenamefont {Huang},
  \citenamefont {Liu}, \citenamefont {Wang}, \citenamefont {Su}, \citenamefont
  {Liu}, \citenamefont {Shi}, \citenamefont {Gao}, \citenamefont {Jiang},
  \citenamefont {Liu}, \citenamefont {Liu}, \citenamefont {Lu}, \citenamefont
  {Yang}, \citenamefont {Zhang}, \citenamefont {Huan}, \citenamefont {Xia},
  \citenamefont {Wang}, \citenamefont {Wu}, \citenamefont {Wang}, \citenamefont
  {Yu}, \citenamefont {Huang}, \citenamefont {Qiao}, \citenamefont {Li},
  \citenamefont {Zhao}, \citenamefont {Guo}, \citenamefont {Li},\ and\
  \citenamefont {Shen}}]{Huang2020}%
  \BibitemOpen
  \bibfield  {author} {\bibinfo {author} {\bibfnamefont {Z.}~\bibnamefont
  {Huang}}, \bibinfo {author} {\bibfnamefont {W.~L.}\ \bibnamefont {Liu}},
  \bibinfo {author} {\bibfnamefont {H.~Y.}\ \bibnamefont {Wang}}, \bibinfo
  {author} {\bibfnamefont {Y.~L.}\ \bibnamefont {Su}}, \bibinfo {author}
  {\bibfnamefont {Z.~T.}\ \bibnamefont {Liu}}, \bibinfo {author} {\bibfnamefont
  {X.~B.}\ \bibnamefont {Shi}}, \bibinfo {author} {\bibfnamefont {S.~Y.}\
  \bibnamefont {Gao}}, \bibinfo {author} {\bibfnamefont {Z.~C.}\ \bibnamefont
  {Jiang}}, \bibinfo {author} {\bibfnamefont {Z.~H.}\ \bibnamefont {Liu}},
  \bibinfo {author} {\bibfnamefont {J.~S.}\ \bibnamefont {Liu}}, \bibinfo
  {author} {\bibfnamefont {X.~L.}\ \bibnamefont {Lu}}, \bibinfo {author}
  {\bibfnamefont {Y.~C.}\ \bibnamefont {Yang}}, \bibinfo {author}
  {\bibfnamefont {J.~X.}\ \bibnamefont {Zhang}}, \bibinfo {author}
  {\bibfnamefont {S.~C.}\ \bibnamefont {Huan}}, \bibinfo {author}
  {\bibfnamefont {W.}~\bibnamefont {Xia}}, \bibinfo {author} {\bibfnamefont
  {J.~H.}\ \bibnamefont {Wang}}, \bibinfo {author} {\bibfnamefont {Y.~S.}\
  \bibnamefont {Wu}}, \bibinfo {author} {\bibfnamefont {X.}~\bibnamefont
  {Wang}}, \bibinfo {author} {\bibfnamefont {N.}~\bibnamefont {Yu}}, \bibinfo
  {author} {\bibfnamefont {Y.~B.}\ \bibnamefont {Huang}}, \bibinfo {author}
  {\bibfnamefont {S.}~\bibnamefont {Qiao}}, \bibinfo {author} {\bibfnamefont
  {J.}~\bibnamefont {Li}}, \bibinfo {author} {\bibfnamefont {W.~W.}\
  \bibnamefont {Zhao}}, \bibinfo {author} {\bibfnamefont {Y.~F.}\ \bibnamefont
  {Guo}}, \bibinfo {author} {\bibfnamefont {G.}~\bibnamefont {Li}},\ and\
  \bibinfo {author} {\bibfnamefont {D.~W.}\ \bibnamefont {Shen}},\ }\href@noop
  {} {\bibinfo {title} {Dual topological superconducting states in the layered
  titanium-based oxypnictide superconductor \ce{BaTi2Sb2O}}} (\bibinfo {year}
  {2020}),\ \Eprint {https://arxiv.org/abs/2009.06805} {arXiv:2009.06805}
  \BibitemShut {NoStop}%
\bibitem [{\citenamefont {Mazin}\ \emph {et~al.}(2008)\citenamefont {Mazin},
  \citenamefont {Johannes}, \citenamefont {Boeri}, \citenamefont {Koepernik},\
  and\ \citenamefont {Singh}}]{Mazin2008}%
  \BibitemOpen
  \bibfield  {author} {\bibinfo {author} {\bibfnamefont {I.~I.}\ \bibnamefont
  {Mazin}}, \bibinfo {author} {\bibfnamefont {M.~D.}\ \bibnamefont {Johannes}},
  \bibinfo {author} {\bibfnamefont {L.}~\bibnamefont {Boeri}}, \bibinfo
  {author} {\bibfnamefont {K.}~\bibnamefont {Koepernik}},\ and\ \bibinfo
  {author} {\bibfnamefont {D.~J.}\ \bibnamefont {Singh}},\ }\bibfield  {title}
  {\bibinfo {title} {Problems with reconciling density functional theory
  calculations with experiment in ferropnictides},\ }\href
  {https://doi.org/10.1103/PhysRevB.78.085104} {\bibfield  {journal} {\bibinfo
  {journal} {Phys. Rev. B}\ }\textbf {\bibinfo {volume} {78}},\ \bibinfo
  {pages} {085104} (\bibinfo {year} {2008})}\BibitemShut {NoStop}%
\bibitem [{\citenamefont {Graser}\ \emph {et~al.}(2009)\citenamefont {Graser},
  \citenamefont {Maier}, \citenamefont {Hirschfeld},\ and\ \citenamefont
  {Scalapino}}]{Graser2009}%
  \BibitemOpen
  \bibfield  {author} {\bibinfo {author} {\bibfnamefont {S.}~\bibnamefont
  {Graser}}, \bibinfo {author} {\bibfnamefont {T.~A.}\ \bibnamefont {Maier}},
  \bibinfo {author} {\bibfnamefont {P.~J.}\ \bibnamefont {Hirschfeld}},\ and\
  \bibinfo {author} {\bibfnamefont {D.~J.}\ \bibnamefont {Scalapino}},\
  }\bibfield  {title} {\bibinfo {title} {Near-degeneracy of several pairing
  channels in multiorbital models for the fe pnictides},\ }\href
  {https://doi.org/10.1088/1367-2630/11/2/025016} {\bibfield  {journal}
  {\bibinfo  {journal} {New J. Phys.}\ }\textbf {\bibinfo {volume} {11}},\
  \bibinfo {pages} {025016} (\bibinfo {year} {2009})}\BibitemShut {NoStop}%
\end{thebibliography}

\providecommand{\noopsort}[1]{}\providecommand{\singleletter}[1]{#1}%

\end{document}